\documentclass[preprint2]{aastex}

\slugcomment{To appear in Astronomical Journal}

\shorttitle{\hi~properties of clusters of galaxies}
\shortauthors{Venkatapathy et al.}

\newcommand{\hi}{H{\sc i}}

\newcommand{\prim}{$^{\prime}$}
\newcommand{\prin}{$^{\prime\prime}$}
\newcommand{\aprox}{${\sim}$}

\newcommand{\kms}{~km~s$^{-1}$}

\newcommand{\degree}{$^{\circ}$}

\newcommand{\msolar}{M$_{\odot}$}

\newcommand{\por}{$\times$}
\newcommand{\atn}{${\alpha}_{An}$}
\newcommand{\atf}{${\alpha}_{A3}$}
\begin{document}


\title{Environmental effects on galaxy evolution. II:
quantifying the tidal features in NIR-images of the cluster Abell\,85}


\author{Y. Venkatapathy,\altaffilmark{1}}
\affil{Departamento de Astronom\'\i a, Universidad de Guanajuato, Apdo. Postal 144, Guanajuato 36000, Mexico}

\author{H. Bravo-Alfaro,\altaffilmark{1,2}}
\affil{Departamento de Astronom\'\i a, Universidad de Guanajuato, Apdo. Postal 144, Guanajuato 36000, 
Mexico}
\affil{Institut d'Astrophysique de Paris, CNRS, UMR 7095, Universit\'e Pierre et Marie Curie, 
98bis Bd Arago, 75014 Paris, France }
\email{hector@astro.ugto.mx}

\author{Y. D. Mayya\altaffilmark{3}}
\affil{Instituto de Astrof\1sica, \'Optica y Electr\'onica. Luis Enrique Erro \# 1, Puebla 72840, M\'exico}

\author{C. Lobo\altaffilmark{4,5}}
\affil{Instituto de Astrof\'\i sica e Ci\^encias do Espa\c co, Universidade do Porto, CAUP, 
Rua das Estrelas, PT4150-762 Porto, Portugal}
\affil{Departamento de F\'\i sica e Astronomia, Faculdade de Ci\^encias, Universidade do Porto, Rua do
Campo Alegre 687, PT4169-007 Porto, Portugal}

\author{F. Durret\altaffilmark{2}}
\affil{Institut d'Astrophysique de Paris, CNRS, UMR 7095, Universit\'e Pierre et Marie Curie, 
98bis Bd Arago, 75014 Paris, France }

\author{V. Gamez\altaffilmark{1}, M. Valerdi\altaffilmark{1}, A. P. 
Granados-Contreras\altaffilmark{1}, and F. Navarro-Poupard\altaffilmark{1}}
\affil{Departamento de Astronomia, Universidad de Guanajuato, Apdo. Postal 144, Guanajuato 36000, Mexico}


\altaffiltext{1}{Departamento de Astronomia, Universidad de Guanajuato.
Apdo. Postal 144, Guanajuato 36000. Mexico}
\altaffiltext{2}{Institut d'Astrophysique de Paris, CNRS, UMR 7095, Universit\'e Pierre et Marie Curie, 
98bis Bd Arago, 75014 Paris, France}
\altaffiltext{3}{Instituto de Astrof\1sica, \'Optica y Electr\'onica. Luis Enrique Erro \# 1, 
Puebla 72840, M\'exico}
\altaffiltext{4}{Centro de Astrof\1sica da Universidade do Porto, Rua das Estrelas, 4150-762 Porto, 
Portugal}


\begin{abstract}

This work is part of a series of papers devoted to investigate the 
evolution of cluster galaxies during their infall. 
In the present article we imaged in NIR a selected sample of 
galaxies throughout the massive cluster Abell\,85 (z\,=\,0.055). 
We obtained (JHK') photometry for 68 objects,
{\bf reaching \aprox1 mag\,arcsec$^{-2}$ deeper than 2MASS}.
We use these images to unveil asymmetries in the outskirts
of a sample of bright galaxies and develop a new asymmetry 
index, \atn, which allows to quantify the degree of disruption 
{\bf by the relative area occupied by the tidal features}
on the plane of the sky.  
We measure the asymmetries {\bf for a subsample of 41 large area
objects finding} clear asymmetries in ten galaxies, most
of them being in groups and pairs projected at different clustercentric
distances, some of them located beyond $R_{500}$.  Combining  
information on the \hi-gas content of blue galaxies and the 
distribution of sub-structures across Abell\,85, 
with the present NIR asymmetry analysis, we obtain a very powerful 
tool to confirm that 
tidal mechanisms are indeed present and are currently affecting
a fraction of galaxies in Abell\,85.
However, when comparing our deep NIR images with UV-blue images 
of two very disrupted ($jellyfish$) galaxies in this cluster, 
we discard the presence of tidal interactions down to our 
detection limit. Our results suggest that ram-pressure stripping is 
at the origin of such spectacular disruptions. 
We conclude that across a complex cluster like Abell\,85, environment 
mechanisms, both gravitational and hydrodynamical, are playing an 
active role in driving galaxy evolution.

\end{abstract}



\section{Introduction} \label{intro}

Since several decades, many efforts have been devoted to understand the
origin of the density morphology relation  
\citep[see e.g.][and references therein]{dress80}.
The fact that in the nearby universe spiral galaxies are systematically 
less abundant in the central cluster regions, compared with the field,
constitutes important evidence that environment has been 
playing an important role in galaxy evolution at least since $ z \sim$ 
0.5 \citep{Lewis02,koop04,BoGa06,Bam09,Jaffe11, Jaffe16}.
Inversely, this remarkable absence of spirals near the cluster cores is
accompanied by a growing population of lenticulars 
towards high density regions. This raises the question of whether a 
large fraction of spirals is being transformed into S0's 
during their infall towards galaxy clusters \citep{Kodama01,Erwin12,Rawle13}.  
Some authors propose that two types
of lenticulars could exist, one of them corresponding 
to the {\it processed} spiral galaxies  \citep{Bedregal06,Barway07,Calvi12}.

There is indisputable evidence for
cluster environment effects on individual galaxies, 
but determining which are the main physical processes 
driving galaxy evolution is still a matter of debate. 
Such mechanisms are classified in two types: the
hydrodynamic mechanisms exerted by the hot intra cluster 
medium (ICM) e.g. ram pressure stripping \citep{Gunn72}
and gravitational processes. The latter include both galaxy-galaxy
and/or galaxy-cluster interactions \citep{Merritt83,Byrd90,Moore96}.
Presently, it is currently accepted that more than one single
mechanism must be at work, specially on the galaxies undergoing strong
morphological transformations during their infall onto the cluster
\citep{Cortese07,Yoshida12,Ebeling14,McPartland16}.

While imaging the HI in late-types is a direct tool to study 
ram-presure stripping \citep{Bravo00,Bravo01,Poggianti01,Kenney04,
  Crowl05,Chung07,Chung09,Scott10,Scott12,Jaffe16},
tracing the tidal features is not always straightforward,
observationally speaking.
First, such structures show up (most times) at low 
surface brightness. Second, the old stellar population is 
a very good tracer of gravitational tidal structures
\citep{Plauchu10}
and these stars are better seen in NIR; in this band 
the contamination produced by star forming regions (at
least in the case of spirals) is lower than in the
optical bands.  Therefore deep NIR 
imaging is best suited to study tidal disruptions
\citep[e.g. WINGS,][]{Valentinuzzi09}.  Traces of old 
stars (several gigayears old) found along tidal 
structures constitute the smoking gun of gravitational
interactions. Old stars could only be
teared up from the galaxy disk by tidal interactions,
while young stars could be formed in situ from ram-pressure  
stripped gas.
However, observing in the NIR raises a number of difficulties,
starting with the fact that relatively complex techniques
are needed to observe and reduce data in the J, H, K bands.
Using the available 2MASS images is not an option to tackle 
this issue when dealing with objects at $z$\aprox0.01 (and beyond),
because these images are not deep enough to unveil the low
surface brightness features (see Sect. \ref{phot}).

To complicate the issue, a direct comparison between different 
surveys is rarely straightforward, either because the observed 
samples are different (in morphology, environment or redshift) 
or because the method to quantify the tidal features are
not the same \citep[see e.g.][and references therein]{Adams12}.
In this respect, \citet{Holwerda14} 
reviewed the methods presently available to determine
galaxy morphologies and tidal features.
They reported several 
criteria to identify disturbed galaxies, involving 
parameters like the flux 2nd-order moment \citep{Lotz04}, 
the CAS system \citep{conselice03},
and the Gini index \citep{Abraham03}.

With the aim of quantifying the role played by tidal interactions
in the evolution of galaxies in nearby clusters, we develop a 
new asymmetry index which is well suited to measure low surface 
brightness asymmetries in the outskirts of galaxies.  Our method 
is applied here to a few dozens of member galaxies of Abell\,85
(z\,=\,0.055), a complex system where tidal mechanisms are expected 
to be significant.  This cluster has a large
set of observational data, going from X-ray \citep{Ichinohe15}, 
VLA-HI data (Bravo-Alfaro, et al. in prep.),
to optical imaging and spectroscopy.  The studies of the 
substructures of A\,85 \citep{Durretb,Bravo09,Yu16}, provide 
useful information to correlate with gravitational pre-processing.
This paper constitutes a step further in a broader study on
galaxy evolution in clusters; our previous work on A\,85
\citep{Bravo09} tackled the corrrelation between the 
\hi-content of late-types and their position within
different substructures throughout the cluster. 
The following papers of this series will be devoted to
large field coverage of several nearby systems (including A\,85,
A\,496 and A\,2670), both in NIR and \hi, in order to study the
evolution of galaxies on statistical basis.

The present paper is organized as follows. In Sect.~\ref{obs} we describe 
our survey, the NIR-observing strategy, and the flux calibration. 
We provide a NIR (J,H,K') catalog for the sampled galaxies.
In Sect. \ref{results} we describe our method to measure the asymmetry 
features.   We compare our asymmetry index with other tools 
currently available in the literature. 
In Sect. \ref{discussion} we discuss the fraction of
galaxies showing asymmetries and their positions across A\,85.
We present a global view of the cluster confirming the presence of
some physical pairs and groups of galaxies, taking into account 
the tidal interactions we unveil in the NIR.
In Sect. \ref{selected_fields} we describe the most interesting
cases of individual galaxies in selected fields, including 
three very disrupted objects (two of them being classified as
$jellyfish$ galaxies).
Sect. \ref{concl} provides a summary and our main conclusions.

Throughout this paper we assume $\Omega_M$\,=\,0.3, 
$\Omega_\Lambda$\,=\,0.7, and H$_0$\,=\,75 km s$^{-1}$\,Mpc$^{-1}$.
In this cosmology, 10$^\prime$\, are equivalent to \,0.6\,Mpc at the 
distance of Abell\,85.

\section{Observations and data reduction} \label{obs}

\subsection{The sample}\label{sample}

Fig. \ref{A85-fields} shows the location of the 26 fields observed
throughout A\,85.  These fields were chosen under several criteria
(see Column 7 of Table \ref{fld}).  
First, we targeted the ten \hi-detections (excluding two marginal ones)
reported by \citet{Bravo09} (hereafter BA09).   This is
with the aim of studying their evolutionary stage while moving
towards/across the cluster.  Second, after visual inspection, we
selected the fields displaying (at least in projection) pairs 
or groups of bright galaxies.  These are places where we expect 
to see tidal features at different degrees.  Another field 
was devoted to the cD, A85[DFL98]242, and one more to
a galaxy showing some asymmetries (A85[DFL98]276), but being 
apparently isolated (under projection and velocity criteria),
see Table~\ref{fld}.

Our selected fields include the brightest galaxies in A\,85.
With a few exceptions, all these objects are members of A\,85,
following the membership (position-velocity) criteria given by BA09.
The observed sample is complete up to $B_{J} = 16$ (following
the SuperCOSMOS database),  for the redshift range of the 
cluster and within a region going from 
00h 40m 30s to 00h 44m 00s in R.A., and from -08\degree\,45\prim\, 00\prin\, 
to -10\degree\, 05\prim\, 00\prin\, in declination.  This sample
is devoted to obtain a first insight on the presence of tidally
disrupted galaxies in A\,85, and to quantify their degree of 
asymmetry. In total, we obtained NIR magnitudes for 68 galaxies, 
projected inside a radius of 1\degree\, from the cluster center,
which we take as coincident with the position of the cD galaxy.
Table \ref{opt} gives the optical parameters of the observed
objects.

\subsection{Image acquisition and processing}\label{improcess}

We selected 26 fields in the A\,85 galaxy cluster to be observed
in the NIR bands JHK' (1.28, 1.67, 2.12~$\mu$m). The K' filter is 
available at the OAN instead of the K-band one; their effective 
wavelengths are the same, i.e. $2.12\mu$m.
All our images were obtained between 2006 and 2011 at the 2.1m
telescope of the National Astronomical Observatory (OAN), in San Pedro
Martir, Mexico.  We used the NIR camera CAMILA \citep{Cruz94}, equipped
with a 256\por256 pixels NICMOS3 detector array. The image scale is
0.85\prin\,pixel$^{-1}$, and the field of view is 3.6\prim\por
3.6\prim. Some optical vignetting reduced the useful {\sl fov} to
3.0\prim.  The seeing during our observing runs varied between
2.0\prin and 2.5\prin.

Due to the high sky brightness and variability, seen in the NIR, 
we chose a ``telescope chop'' strategy, in order to properly
scan the sky background in every band. Typically, our galaxies do not 
extend over a large fraction of the detector, most of them having a
major axis well below one arcminute. So instead of the  "on/off
target" strategy, we offset the pointing-center by less than one
arcminute between exposures.  With this technique we manage to keep
the targets on different zones of the CCD, distributed along the
four detector quadrants, thus saving observing time. 

The linearity range of the detector constrained us to apply short
individual sub-exposures in order to avoid saturation.  These 
limits are typically 30s, 20s, 5s, for J, H and K', respectively. 
We made sequences of 9 pointings of 60s each, splitting the 60s,
in order to avoid saturation, into 2x30s, 3x20s and 12x5s, 
depending on the waveband. We applied the mentioned offsets 
between pointings, and we repeated the sequences until
reaching total integration times in
the range 1600s-3800s (see Table \ref{fld}).
With this strategy, the median 
average of the nine frames provides a good sky image, where the
cosmic rays, the stars and the galaxies themselves, have been removed.

The image processing and calibration were performed
using IRAF.\footnote{
IRAF is distributed by the
National Optical Astronomy Observatory, which is operated by the
Association of Universities for Research in Astronomy, Inc., under
cooperative agreement with the National Science Foundation.}
We followed standard procedures for data reduction,
following \citet{Barway05} and \citet{Romano08}.
For the flat-fielding we applied the twilight sky method and obtained
two different sets of images, those with high count levels (``bright flats'') 
and those with low count levels (``dark flats''). We combined the dark and
bright flats separately, and then subtracted the dark flat from 
the bright one.  The resulting frame was normalized by its mean value, 
and this master flat was used for general flat fielding. This procedure
was repeated for each observed band.

A key step is the sky subtraction.  We combined all the frames, within 
a nine-image sequence (see above), by using the median 
criteria.  The resultant frame constitutes a good sky image, which is
subtracted from each individual image of the corresponding sequence.
The resulting sky-free images were aligned to a common coordinate
system by using stars appearing in all frames. Finally, these images
were averaged, delivering a final, cleaned image for each band.
All the procedure described above was carried out by applying the 
script $CAMILA$, developed by one of us (YDM).  As a last step, 
we carried out the corresponding astrometry, by matching the galaxy 
coordinates using the 2MASS catalog from the NASA/IPAC Infrared 
Science Archive (IRSA).

\subsection{Flux calibration and photometry} \label{phot}

We observed a set of photometric standard stars in order to
carry out the flux calibration. Here, the strategy for image 
reduction followed the procedure described in Sect. \ref{improcess}.
The only difference was the application of shorter 
integration times (a few seconds) for
the standard stars, as their magnitudes are much
brighter than those of our science targets.  The observed
standards (FS 101, FS 104, FS110, FS111,
FS112, FS119, FS150, FS154)
were selected from the  \citet{Persson98}  and \citet{Hawarden01}
catalogs.   We observed several standards during each night,
under different airmasses, in order to improve the accuracy of
our zero-points (ZP).  We solved the following equation to calculate 
the ZPs in each band:

\begin{equation}
m_{\lambda} = -2.5 log R_{\lambda} + ZP_{\lambda}
\end{equation}

\noindent
where $R_{\lambda} $ represents the instrumental counts  
and $ZP_{\lambda}$ is the zero-point constant, in each filter.
NIR zero-points are found to be stable, not only during 
a single night, but over several nights through the whole 
observing run.  For each night we have estimated the 
known magnitude of one standard star, by using the ZP of 
other standards observed along the same night.  The 
magnitudes obtained this way, matched very well with 
each other.  In the end we average the individual $ZP$
coming from different stars, ensuring that the ZP 
value is correct.

We obtained aperture magnitudes (14\prin\, in diameter) for galaxies in our sample
as this allows direct comparison with published NIR catalogs. 
We measured these magnitudes with
SExtractor \citep{Bertin96}.  We first ran SExtractor in single-image mode
for the J-band, which has the highest S/N ratio. We then ran  SExtractor again
in dual mode, with the J-band frame as the reference, and we computed the
magnitudes of H and K' bands. This
procedure ensured that the measurements of each galaxy, in all three
filters, were done exactly over the same pixels.
In the end, from the 71 objects targeted in Table \ref{opt}.
we obtained good quality images and magnitudes for 68 galaxies
(see Table \ref{magn}).  Three of 
the objects previously detected in HI ([SDL98]3114,
and [DFL98]323/461, see BA09) are faint objects and have very 
blue colors, therefore, our NIR images 
did not provide enough signal to obtain accurate magnitudes.
These objects are not included in our further analysis.

\begin{figure*}
\epsscale{1.0}
\plotone{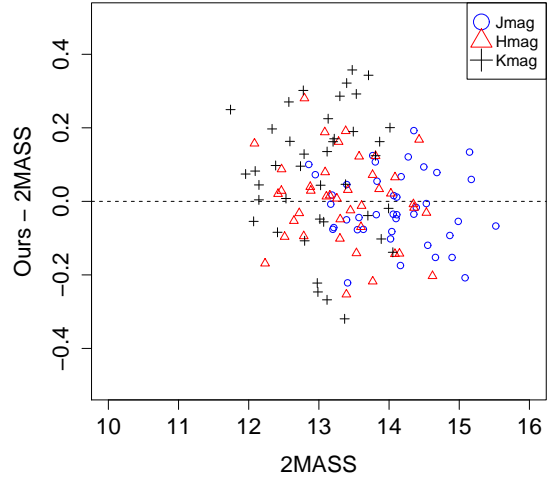}
\caption{Comparison of 2MASS magnitudes with the NIR values obtained in this work ($Ours$). 
The X-axis displays 2MASS magnitudes and the Y-axis indicates the difference 
$Ours - 2MASS$.  {\bf A slight trend is observed on our K' magnitudes being -in 
average- larger than 2MASS'.  This is not shown by the other two bands (see text).}
\label{2MASS_comp}}
\end{figure*}

Forty-one galaxies of the 68 in our photometric sample have NIR 
magnitudes in the 2MASS catalog. We compared our 14\prin\,
aperture magnitudes, shown in Table \ref{magn}, with those 
published by 2MASS.   This allows to test the quality of 
our imaging procedure and the accuracy of our photometric data. 
We have found a good match between both catalogs 
(see Fig. \ref{2MASS_comp}).  As expected, we obtained 
a slightly larger deviation in the K'-band, as the noise is higher 
at this frequency.  
{\bf
Our magnitudes in this band display a slight trend being, in
average, larger than 2MASS. The most likely 
explanation is linked to the (rather long) age of the
NIR camera at the time of our observing runs; the detector 
and/or other optical parts of the instrument could lose 
sensitivity after decades of service. If so, this could affect in first 
term the K' band, rather than J, H, where the effect is not seen 
(Fig. \ref{2MASS_comp}). Nevertheless, this bias is not significant:
the average differences between our magnitudes and those from 
2MASS (in absolute values), and corresponding standard deviations,
are 0.08$\pm$0.05 for J, 0.10$\pm$0.08 in H, and 0.16$\pm$0.10 in K'.
The first advantage of our survey, compared with 2MASS (within 
the observed area), is the higher number of galaxies with 
reported magnitudes.  And second, our frames are deeper by 
(roughly) one magnitude arcsec$^{-2}$ when compared with 2MASS.  
Our survey reached, on average, the following 1-sigma 
background noise and corresponding errors: 
22.40$\pm$0.06, 21.20$\pm$0.07, and 20.30$\pm$0.06
mag~arcsec$^{-2}$, in J, H, and K', respectively.  
The corresponding 2MASS values are 21.4, 20.6, and
20.0 \citep{Jarrett03}. 
}

\begin{figure*}
\epsscale{1.0}
\plotone{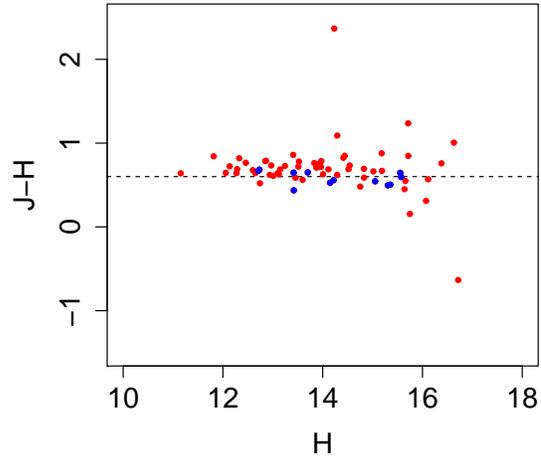}
\caption{Color-magnitude plot {\it (J-H) vs H} of isophotal magnitudes of 
the 68 galaxies observed in this work. Blue dots correspond to blue
galaxies, as reported by BA09. The red dots indicate early-type 
galaxies.  The dotted line at (J-H) = 0.6 is a reference for the red sequence 
for NIR data. 
\label{color-mag}}
\end{figure*}

In order to illustrate the photometric properties of our sample 
we display a color-magnitude diagram (J-H) vs J (Fig. \ref{color-mag}), 
based on our total isophotal magnitudes. These values were obtained 
with SExtractor, with a detection threshold of 1$\sigma$ measured 
on the background.  We plot, as a reference, a red 
sequence (at J-H\,=\,0.6) derived using 2MASS data (Caretta 2015, priv. 
comm.)  This figure shows that our sample is rather dominated by red 
objects. Two galaxies appear with extreme colors in this plot, and should
be taken with caution. One of them is displaying an abnormal 
red color (A85[SDG98]1951), probably because of contamination of
the neighbor cD halo (see Sect. \ref{selected_fields}). Another galaxy
(A85[SDG98]2260), appears with an extremely blue color (bottom-right
corner of Fig. \ref{color-mag}); this object is lying on the very edge 
of the corresponding field, which could have affected its photometry.

\section{Measuring the asymmetry features} \label{results}

\subsection{The asymmetry index \atn} \label{index_def}

The main goal of this work is to detect and quantify asymmetry
features in galaxies produced through tidal
interactions.  With this aim we apply an asymmetry analysis
which is focused on the old stellar morphology drawn by NIR 
images.  Measuring asymmetries has proven to be useful with images 
at different wavelengths, including optical and \hi. 
This work is intended to be a first approach 
to measure the role played by gravitational mechanisms in the 
evolution of galaxies in A\,85, within its
middle and high density regions.  

Visual inspection remains as one of the best suited techniques 
to classify galaxies \citep{McIntosh04,Mihos05}. However, considering 
the huge amount of data available nowadays, this method is very limited.
Furthermore, visual classification does not provide quantitative
information, for instance, about the degree of disruption a galaxy
is undergoing, thus reducing the possibility of any statistical study.
This raises the importance of methods that quantify the
morphological properties of galaxies, as they allow to correlate
those properties with environment conditions and, in the end, 
to shed light on the physics driving galaxy evolution.

Our strategy to measure galaxy asymmetries
was the following. 
{\bf 
First, we select within our sample those
galaxies having angular dimensions above a certain threshold
in order to keep only those objects with enough data points.
This subsample consists of the 41 galaxies having major 
axis $a \geqq 15$\prin\, (or \aprox18 pixels),
and minor axis $b \geqq 7.5$\prin, (\aprox9 pixels).
As our uncertainty is dominated by the seeing (i.e.
2.5\prin\, or 3 pixels), our criteria implies that 
we keep a maximum linear uncertainty of 16\%\, 
on $a$, and 30\%\, on $b$. Propagating these errors when 
calculating the area of the galaxies, we keep an 
uncertainty below the threshold $\leq$33\%.

Next, for the 41 selected galaxies 
}
we generate a 2-D intensity map by applying 
the IRAF task ELLIPSE (STSDAS package). We apply this technique only 
to J-band images, as they have a more homogeneous background and 
higher S/N ratio than the H and K' frames. The ELLIPSE routine,
described by \cite{Jedrzejewski87}, calculates the Fourier series:

\begin{equation}
I(\phi) = I_{0} + \Sigma \  a_{n} \ sin  \ (n\phi) + \Sigma \ b_{n} \ cos  \ (n\phi) 
\label{c}
\end{equation}

\noindent
where $\phi$ is the ellipse eccentric anomaly, $I_{0}$  is the mean 
intensity along the ellipse, and $a_{n}$, $b_{n}$ are harmonic 
amplitudes, along the major and minor axis, respectively.
Typically we start the fit at 2.5\prin\, from the center of
the galaxy, which avoids the bulge (for spirals) and minimizes 
the effects produced by the seeing (see Sect. \ref{improcess}). 
This fitting provides the mean radial light 
distribution and the variation of the three parameters:
center ($x_o, y_o$), ellipticity ($E$), and position angle ($p.a.$), 
as a function of the galaxy radius.
We stop the fitting when reaching isophotes having counts equal to 
three times the standard deviation of the background ($3\sigma_{bg}$), 
equivalent to 
a surface brightness of 21.2 mag arcsec$^{-2}$, in average. For
most of our galaxies, this 
occurs at a radius of \aprox12 pixels, 
equivalent to a linear radius of 10\,kpc from the galaxy center.

We run a second iteration of ELLIPSE; this time we fix the three
parameters (center, $E$, $p.a.$) to values obtained around 6-7\,kpc 
from the galaxy center, with the aim of avoiding the outskirts.  In this 
fashion we get a final intensity profile which is used as input 
of the IRAF task BMODEL.  This task will produce a 2D axy-symmetric 
model of the galaxy, in a frame where the background is defined 
as zero.  Hereafter, this symmetric "clone" 
of the galaxy,  will be named the $bmodel$, which is subtracted from 
the original object, delivering a residual image.  The original
image and the residual one will be named the $observed$ and the 
$residual$ images, respectively.  
During this procedure, the central pixels of the galaxy are actually 
not considered in our analysis, as we are rather interested in
the galaxy outskirts, where we expect less bound material to  
be more easily distorted when tidal effects are exerted on the galaxy.

{\bf 
Based on the residual described above, we define a new asymmetry index, 
named \atn\, ($A$ for "area"; and $n$ makes reference to the 
threshold applied above the background level).  In simple terms, 
our index is described by the following expression:

\begin{equation}
{\alpha}_{An} =  N_{res} (> n*\sigma)~ / ~N_{tot} (> n*\sigma)  
\label{c}
\end{equation}

\noindent
where $ N_{res} (> n*\sigma)$ is the number of pixels measured
upon the residual image having counts above the limit 
$n*\sigma$. $N_{tot} (> n*\sigma)$ is the number of pixels of 
the parent galaxy registering counts above the same cutoff. 
We stress that this limit is the same standard deviation 
of the background ($\sigma_{bg}$) described in the previous 
paragraphs. We keep $\sigma$ in our equations in order
to make the notation simpler.
}

This tool is defined 
to deliver complementary information to that provided, for instance,
by the CAS asymmetry index and by other tools, like the Gini equality 
parameter.   {\bf Our index is devoted to measure how prominent 
(in surface) are the asymmetry features in a galaxy, compared
with the galaxy itself.}  In other words, \atn\,  gives the relative 
$area$ of the features, normalized to their parent galaxy. 
The physical information provided by the area of asymmetry features 
is complementary to the information provided by tools measuring the 
intensity.  Actually, the \atn\, index is intended to resolve
the ambiguity between two galaxies having the same $A$ (CAS) index, 
where one of them has low surface brightness tidal tails, spread 
on a large area, from another galaxy with small bright features.
Distinguishing between the two cases has important physical 
implications, like applying constraints to the age of the event 
being at the origin of the interaction; this could be done
by comparing the observed asymmetry features with current
models of tidal interactions. From the two cases drawn
above, the first event (with more spread features) would be 
expected to be older than the second one, with small and 
bright asymmetries, more embedded in the inner regions of
the parent galaxy.

The index \atn\, is obtained by measuring the surface of the
asymmetry features, in pixels, upon the residual image obtained 
after the subtraction: $observed - bmodel$.  We apply a cutoff in order
to define the borders of the asymmetry features  and to ensure that our 
index is only taking into account the pixels that are brighter 
than our defined threshold. 

As a second step, we estimate the total area covered by the parent galaxy, 
which is measured on the $bmodel$ image. Here, we apply the same criteria
that we used to measure the asymmetry features, i.e. we consider only
those pixels above the defined surface brightness limit, in order
to obtain the total number of pixels covered by the galaxy.
Obtaining the galaxy size
upon the $bmodel$ image, instead of the original one, makes 
the galaxy measurement more homogeneous.  We finally divide the number of 
pixels of the asymmetry features by the corresponding number of pixels 
of the parent galaxy.  Therefore, \atn\, represents the fractional 
surface of the asymmetry features relative to their parent galaxy
(represented by the $bmodel$).  

Throughout this work we apply a $3\sigma$ cut while 
estimating the asymmetries, so hereafter we will often use the more
specific notation, \atf, for the index.  In practice we 
applied the following generalized equation to
estimate the \atn\, index:\\

${\alpha}_{An} =    [ N_{res} (n_{lo}*\sigma) < I < (n_{hi}*\sigma) ] ~~/ $
\begin{equation}
 [ N_{tot} (n_{lo}*\sigma) < I < (n_{hi}*\sigma) ]  
\end{equation}
  
\noindent
where $I$ is the pixel intensity; $(n_{lo}*\sigma)$ represents the 
lower limit mentionned in the previous paragraphs, and $(n_{hi}*\sigma)$
is an upper clipping applied in order to 
discard a few bright (spurious) spikes remaining at the galaxy
center on the residual images.

\subsection{Error sources} \label{index_err}

{\bf 
Fist of all, in the process of building 
the $bmodel$, we are using an elliptical aperture which includes 
the full galaxy and a large fraction of the sky 
background.  In a few cases, a number of objects (stars and/or galaxies) 
can be included within that aperture. When such objects are too 
close to the studied galaxy we applied a mask procedure before
obtaining the $bmodel$.  The value assigned to
the pixels inside the patch is the same than the average background 
sky. In this fashion we avoid any effect on the asymmetry index 
that could be produced by nearby projected objects. 

Other than the problem of having objects too close
to the galaxy under analysis, a number of additional errors
might affect the asymmetry measurements,
}
as reported by several authors \citep[and references 
therein]{Conselice14}; the most important are: 
(a) the correct identification of asymmetries and of the pixels occupied by 
these features; (b) the separation of the background from $signal$ pixels 
(i.e. those belonging to the main galaxy body $and$ those along the 
asymmetry features); and (c) the determination of 
the central pixel of the galaxy.  

We deal with the first  source of error by obtaining an axial symmetric model
of the galaxy (the $bmodel$) as described in the previous section;
the residual image unveils the asymmetry features.  The second 
source of error, the proper discrimination of background, 
was solved by applying a reasonable intensity threshold to the 
selected pixels. This was applied to both sets of pixels, i.e. 
those coming from the asymmetries (on the residual image), and 
those considered as part of the galaxy (on the $bmodel$ frame). 
We applied everywhere $3\sigma_{bg}$, but in the case when
features appear with low surface brightness, the clipping value could 
be adjusted, for instance, to $2.5\sigma_{bg}$.

Concerning the third source of error, the uncertainty on the 
determination of the central galaxy pixel, 
we confirm, as other authors \citep[e.g.][]{Holwerda14}, that this 
constitutes a major source of error.   In order to estimate the effect 
that this uncertainty exerts on the index \atf\,, we collected the
several central pixel values delivered by the task ELLIPSE 
(see Sect. \ref{index_def}), within a box of 3\prin\, around 
the central intensity peak. 
This 3\prin-box coincides with the maximum seeing-value affecting our 
observations.  We computed the index \atf\, taking into account
each one of these center pixels; in the end we defined the very central
coordinates of each galaxy as those producing the minimum asymmetry index.
Finally, we measured the dispersion of the \atf\, 
values obtained in this fashion, as a good 
indicator of the global error. From the sample 
of 41 galaxies we obtained a standard deviation of 0.006.  
Therefore, we settled an uncertainty of 0.01 in \atf\,  
as a realistic (yet conservative) error value for our
asymmetry index.

\subsection{Comparing \atf\, with other methods} \label{index_comp}

As mentioned before, several tools have been proposed to quantify 
asymmetry features by using  strategies similar to ours \citep[][and 
references therein]{Holwerda14}.  
Our index can take values starting from zero, which would correspond
to a perfectly symmetric galaxy. Otherwise, \atf\, takes positive
values: the higher the index, the larger the asymmetry features compared
with the parent galaxy.  For instance, \atf\,=\,1.0 would
represent the case of asymmetry features with a total surface matching
the area covered by the parent galaxy. 
We find systematically  \atf\,$<$\,1.0, even for disrupted objects;
after applying this index to our sub-sample of 41 galaxies (see 
Table~\ref{asym}) we obtained values in the range, 0\,$<$\,\atf\,$\leq$\,0.32.  
After a detailed inspection of galaxies in our
sample we found that a value of \atf\,=\,0.10 clearly
separates symmetric from asymmetric objects. This value corresponds to
features spanning 10\% of the area of the parent galaxy.

In principle, the method to unveil asymmetries based on the subtraction 
of an axi-symmetric model is better suited to analyze early-type
galaxies.  Nevertheless, if the resolution is high enough, 
this method has shown to successfully trace internal structures
in spirals \citep[see e.g.][]{Mayya05},  
 such as bars and rings, that could increase the \atf\ 
index independently of showing (or not) external tidal features. 
In such cases 
we should apply a simple additional step in order to separate 
internal and external asymmetries. 
In this work, given the angular size of our galaxies and the 
data we have, there was no need of applying this last step.

We carried out some comparisons with other techniques of
measuring galaxy asymmetries, in order to test the performance
and degree of confidence of our index. We briefly describe these 
comparisons.

\subsubsection{\atf\, vs visual classification} \label{comp_visual}

\begin{figure*}
\epsscale{0.9}
\plotone{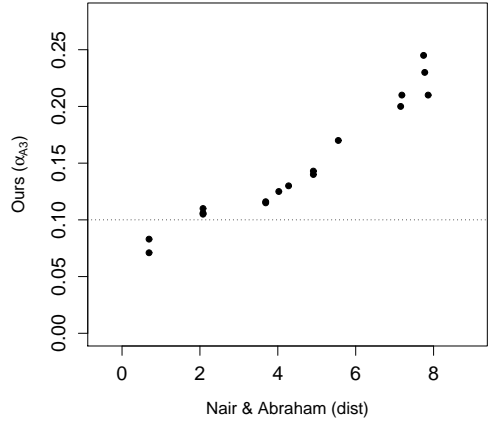}
\caption{Comparison of \atf\, with the visual asymmetry index ($dist$) 
defined by \cite{Nair10}.  Twenty galaxies spanning the full range of 
disruption were chosen. Unperturbed galaxies display index 
values close to zero in both systems, showing that \atf\ successfully
separate symmetric from disrupted objects; the dotted line separates
the former from the later, in our \atf\, system.   A trend is clear where
both indices increase with stronger asymmetry features. 
\label{Nair}}
\end{figure*}

A first test devoted to explore the performance of our asymmetry index 
is the following.  We took into account a subsample of galaxies from
\cite{Nair10}.  
{\bf
These authors carried out a visual morphology classification
for a large sample of galaxies in the range 0.01$<\,z<\,0.1$.
They proposed a discrete, qualitative index ($dist$, 
increasing with the degree and the number of different asymmetries), 
going from fully symmetric up to bridged objects.
We took twenty galaxies from their sample, all being in
the redshift range of Abell\,85 ($z$\aprox0.05), which span the 
whole scale of distortions.  We applied the \atf\, index to 
those objects upon the same g-band (SDSS-DR4) images used by 
\cite{Nair10}.  By comparing these authors' index (see Fig. 
\ref{Nair}) with \atf, we observed that the later is able 
to properly separate the disrupted objects from the symmetric ones:
every galaxy reported by \cite{Nair10} as being unperturbed, 
displays values of \atf\, very close to zero.  In this plot,
objects being reported with important disruptions by \cite{Nair10}, 
would get values above 2.0.  
Fig. \ref{Nair} shows a good trend between the two indices
up to the domain of large asymmetries.  Considering the 
complex way these authors used to define their index (which,
for the twenty selected galaxies, takes values between 1 and 
above 2500) we applied a natural logarithmic scale to their 
original values, so we get a clearer plot.
%
}

\subsubsection{Bmodel vs 180\degree\, rotation} \label{comp_rotat}

Another test to our strategy
consisted in applying a different method to unveil asymmetries.
For example, the asymmetry index $A$, within the CAS system
\citep{conselice03}, measures the asymmetry upon a residual image
which is obtained after rotating a galaxy by 180\degree, then
subtracting this rotated frame from the original image.  The index 
$A$ is based on the integration of the intensities displayed 
by those pixels within the residual features. 

We applied the 180\degree\, rotation method to 
obtain the corresponding residual image, and we calculated the  
\atf\, index for the sample of 41 galaxies 
listed in Table \ref{asym}.
We observed a trend where the $bmodel$ residuals produce
higher values of \atf\, than those coming from
the  180\degree rotation, suggesting that the first method is 
slightly better to unveil outskirt features.
Considering this result, we favor the $bmodel$ strategy over the 
180\degree~rotation. There are additional reasons to favor the
former method; first, the residual delivered after the 180\degree\-rotation 
is very sensitive to the variations 
of the galaxy central pixel, and additional steps are needed to minimize 
this source of error \citep{Conselice14}.  Second, 
the  180\degree rotation method is more sensitive to flocculent and to not-very-regular
spirals; these properties are expected to increase the asymmetry index 
independently of any external disruption \citep{Holwerda14}.  Last, 
but not least, the noise in the  residual image, after rotation, becomes 
very inhomogeneous, complicating the application of any cutoff to compute 
our asymmetry index. In this respect, the advantage of the $bmodel$-subtraction
is that the background of the residual image remains exactly the same than 
in the original image, because the background is defined as zero 
in the $bmodel$ image.  We leave for our forthcoming paper, a direct 
comparison between our index and the $A$ index of the CAS system, as it is 
more convenient to carry out such comparison upon a larger sample of galaxies.

\vspace{0.5cm}

\section{Results and discussion } \label{discussion}

Our asymmetry index, applied in combination with other ones being 
available in the literature, could provide important information 
to restrict the age of tidal interactions at the origin 
of the observed disruptions.  For this, we must compare the
observed galaxies with tidal interaction simulations, taking
into account the time-scales 
delivered by such models \citep[e.g.][]{Lotz04}.  If we only 
consider asymmetries along the outskirts of the galaxies, 
we would expect to find a general trend where recent tidal 
interactions are drawn by stars being projected closer to
their parent galaxy and covering smaller areas than older
events. Tidal interactions, with time, tend to show stars 
spreading through larger regions, making the whole asymmetry
features weaken in surface brightness (the projected 
density of stars will drop as they span through a larger
volume).  In the case a spiral galaxy is affected by a 
tidal encounter, we expect a color gradient to appear;
a recent event will be dominated by blue light
(blue stars are brighter than red ones) and, as time goes on, 
the asymmetry features will become dominated by red light
(as red stars last much longer than blue ones) unless
star formation occurs in situ along the gas tails. 
As a matter of fact, a spiral being recently disrupted 
(not necessarily by tidal interaction) should appear much 
brighter in the UV and blue bands than in the NIR ones. 
We discuss some cases following this trend in 
Sect. \ref{selected_fields}, and we will explore this 
{\it dating} strategy in a forthcoming paper, based on a
larger number of objects.

\subsection{The loci of disturbed galaxies in A\,85 } \label{discuss_1}

\begin{figure*}
\epsscale{2.0}
\plotone{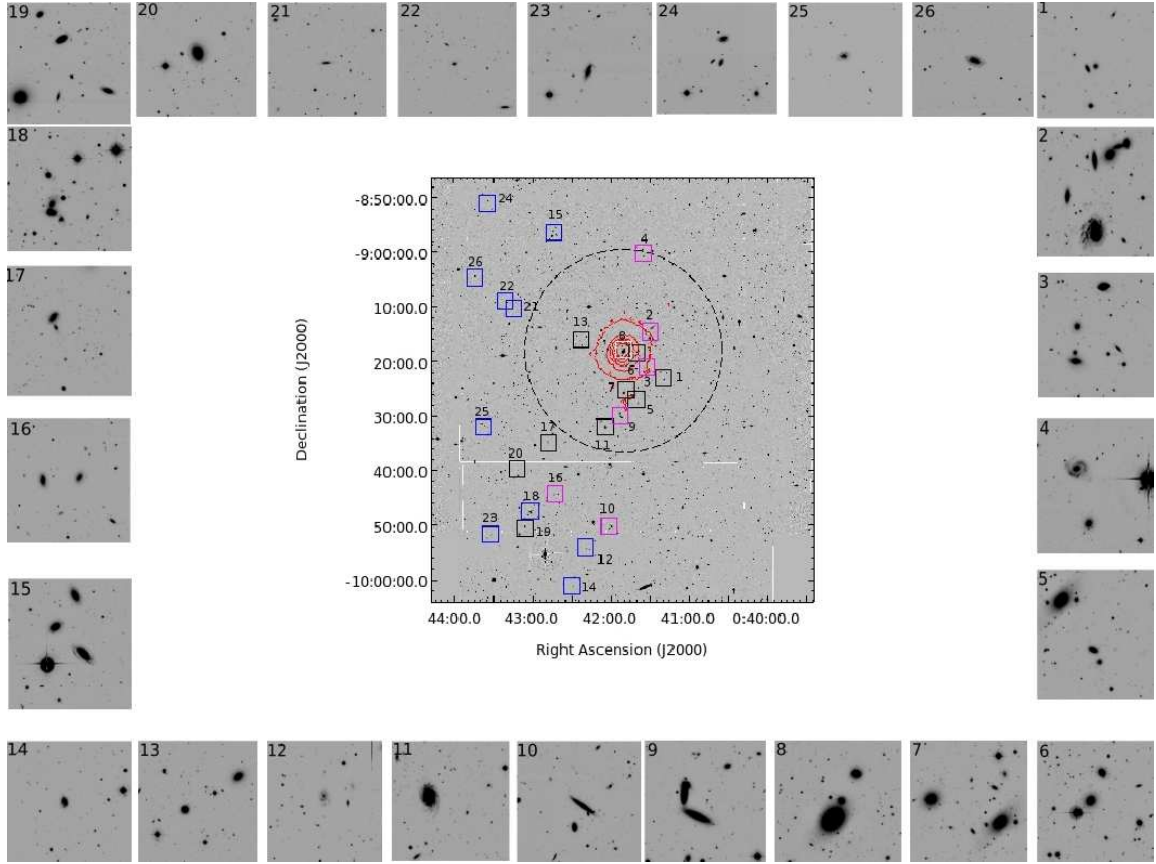}
\caption{Central panel: optical (CFHTg) image of the cluster Abell 85, showing the 
positions and the size of the 26 fields observed in this work. The white square
(Field 8) indicates the position of the cD galaxy. Blue squares correspond to
fields having blue \hi-rich objects. Magenta squares show the location
of fields with asymmetric galaxies. The red contours trace the X-ray emission
(XMM-Newton \citep{Ichinohe15}), and the dotted circle draws the physical
radius R$_{500}$ (\aprox1.2\,Mpc).  A zoom of each field is displayed around the 
central panel; the field number is on the top-left. \label{A85-fields}}
\end{figure*}

\begin{figure*}
\epsscale{1.0}
\plotone{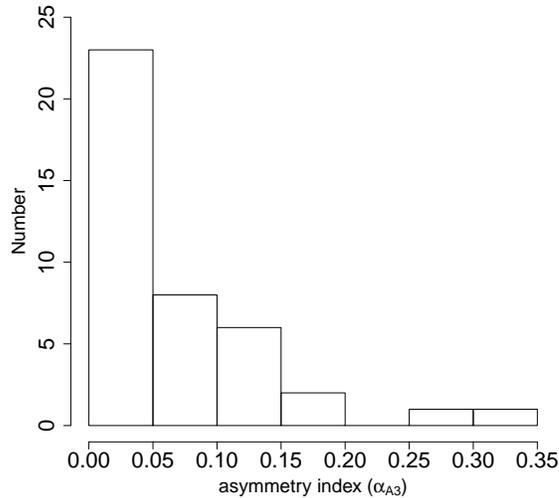}
\caption{Histogram showing the distribution of asymmetry index values for
our sub-sample of 41 galaxies (see Table~\ref{asym}).  Ten of these
objects display significant asymmetries, corresponding to index values \atf\,$>$ 0.1.
\label{histogram}}
\end{figure*}

As expected, combining galaxy positions, radial velocities, 
substructure analysis, and a measurement of 
asymmetries in NIR, constitutes a powerful tool to obtain reliable 
information on the physical mechanisms affecting cluster galaxies.
Furthermore, this strategy allows to confirm (or discard) physical 
pairs and groups, and gives some hints on the degree of 
interaction for those physical pairs/groups. Considering the sample 
of 41 galaxies going through our asymmetry analysis, 
{\bf 
only 10 of them display a significant degree of disruption
(i.e. those having 0.10$<$\atf, see Sect. \ref{comp_visual} 
and Fig. \ref{Nair}).  These asymmetries go from 
mild (\atf\aprox 0.11) to strong (0.14$\leq$\atf$\leq$0.32).  
Fig. \ref{histogram} shows the distribution of asymmetry index 
values of our sub-sample, illustrating that only a fraction
(\aprox 25\%) appears with significant perturbations.
We stress that our sample is coming from selected fields in 
Abell\,85, so this fraction of disturbed objects can be biased. 
We will be tackling this issue, on statistical basis, in our 
forthcoming papers.  

The galaxies with strong asymmetries in the present work
}
are found within six of the 26 observed 
fields. These fields are distributed across A\,85 as follows; two of them 
(Fields No. 2 and 3, see Fig. \ref{A85-fields}) were pointed on possible 
groups of galaxies, showing more than three objects fitting within
a small sky region ($\leq$\,2.5\prim\, equivalent to 150\,kpc).  As seen in 
Fig. \ref{A85-fields}, Fields 2 and 3, are projected onto the cluster
core.
Three other regions with asymmetric galaxies (Fields 4, 9 and 16) 
contain pairs/triplets; the first of these fields is 20\prim\, 
(1.2\,Mpc) north from the cluster center; Field 9 is projected 
onto the {\it South Blob}, \aprox0.75\,Mpc south of the cluster center,
still within the X-ray ICM emission (Fig. \ref{A85-fields}); 
Field 16 is nearly 30\prim\, (\aprox1.8\,Mpc) within the SE sub-cluster
reported by BA09.  Finally, Field 10 is placed at the south outskirts 
of A\,85 (30\prim\, or 
1.8\,Mpc), where a disrupted galaxy appears (intriguingly) isolated.
In the next section we give further details concerning these galaxies 
as well as a couple of other striking objects observed in this work.

\vspace{0.5cm}

\subsection{Comments on selected fields} \label{selected_fields}

\begin{figure*}
\epsscale{2.0}
\plottwo{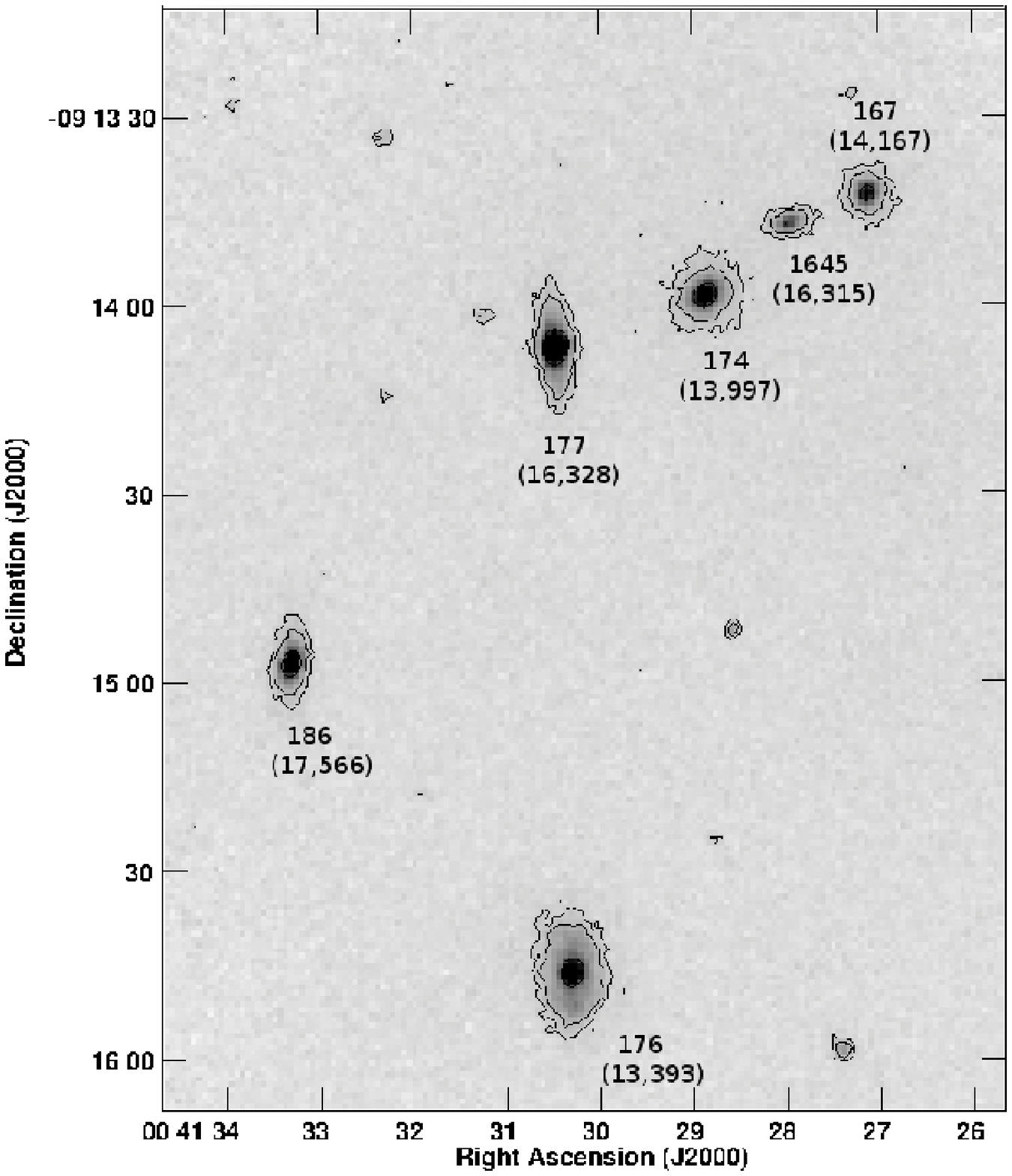}{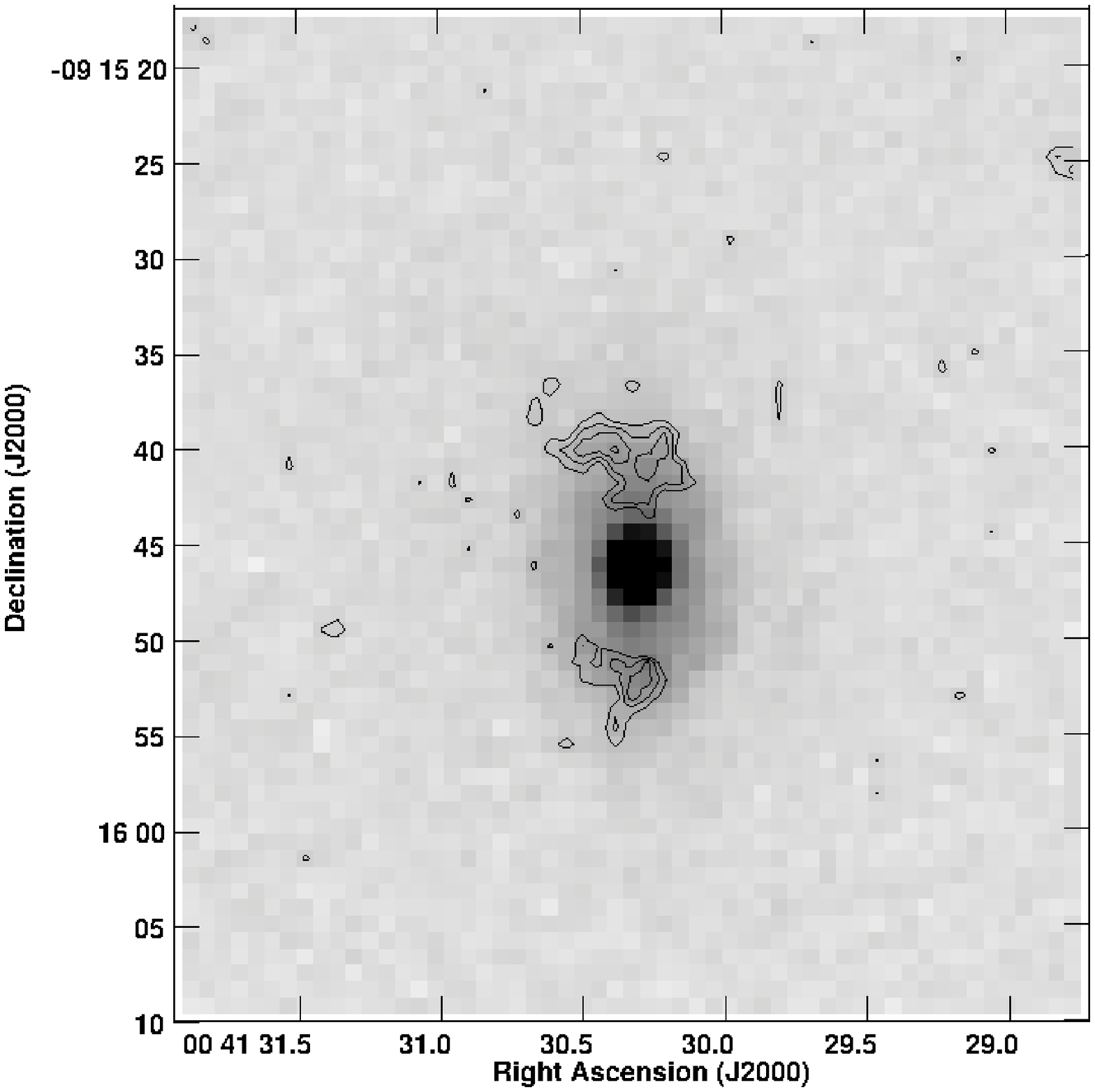}
\caption{Field 2:  A $group$ of galaxies and the $jellyfish$ A85[DFL98]176. Left panel: 
J-band in contours and grey scale. The first contour corresponds to 3.5 times the 
$rms$ background level. The names of the galaxies are given following
Table \ref{opt}, and velocities (in \kms) are given in parentheses.
Right panel: close up of the $jellyfish$ galaxy A85[DFL98]176 (KAZ\,364).
The residual image is shown in white contours, overlaid on the J-band image. 
These contours 
trace slight asymmetries along the northern and southern edges of the disk.  
Compare with the UV and blue images in Fig. \ref{176Jelly}.
\label{176}}
\end{figure*}

\begin{figure*}
\epsscale{0.9}
\plotone{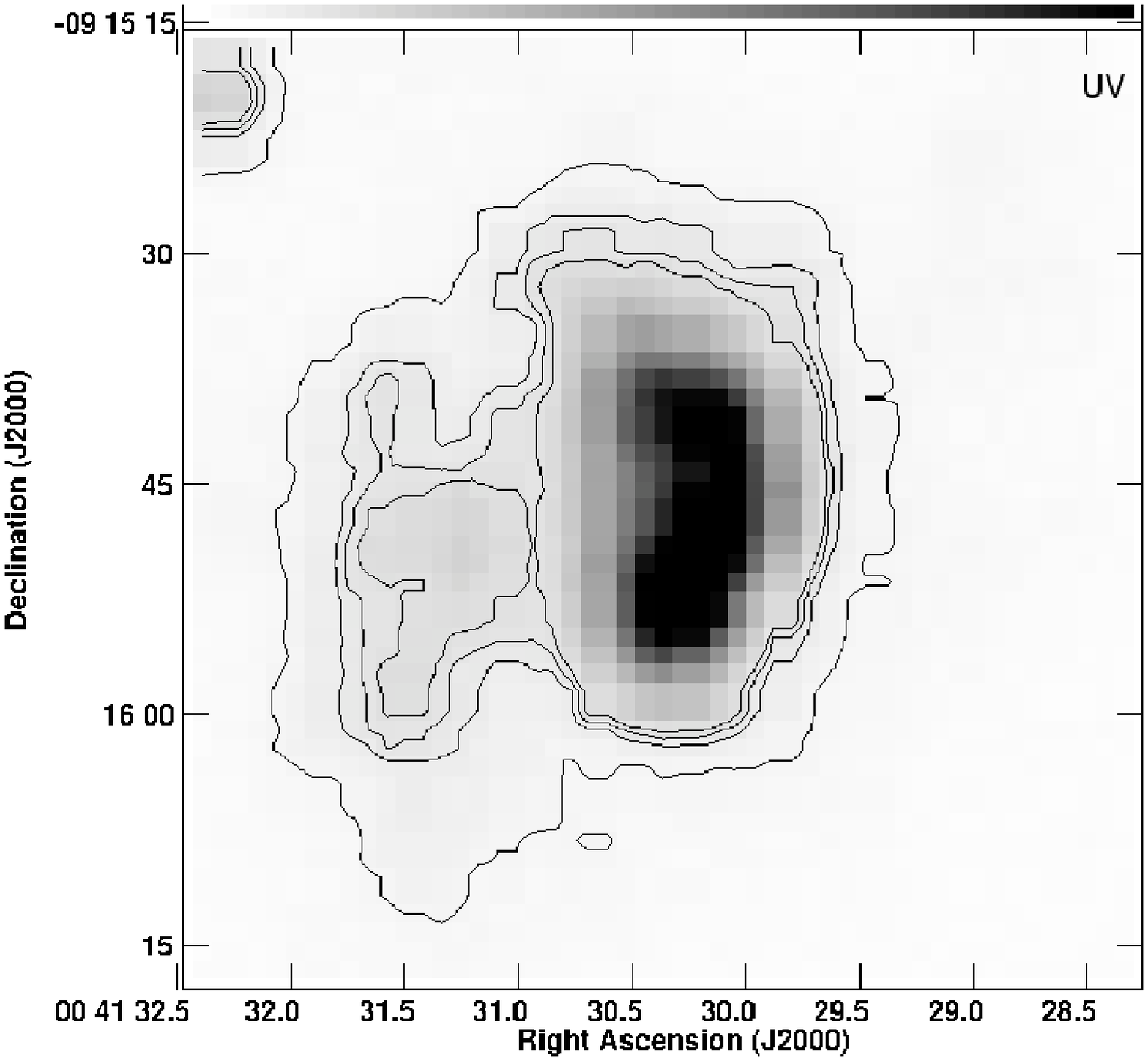}
\plotone{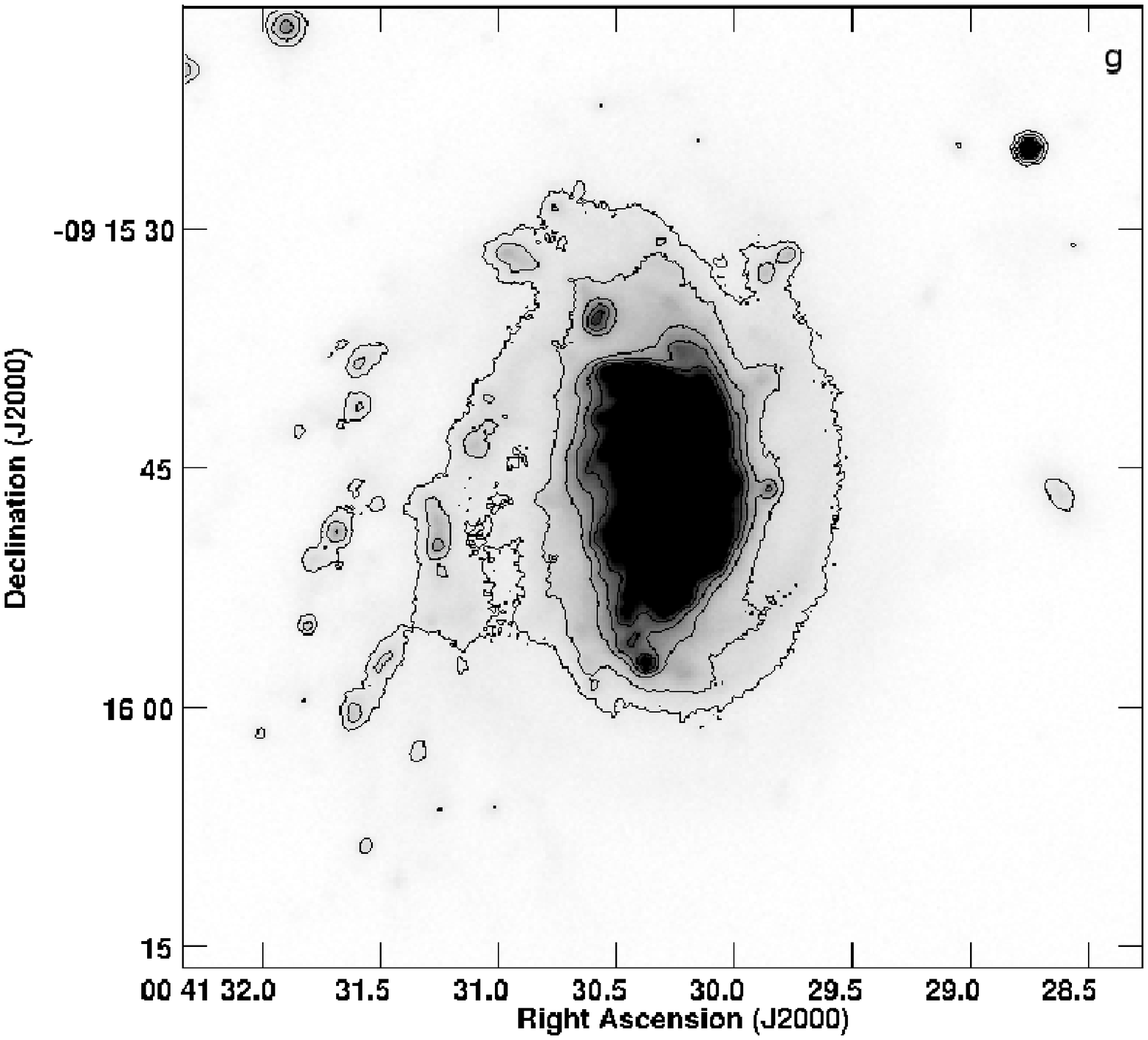}
\plotone{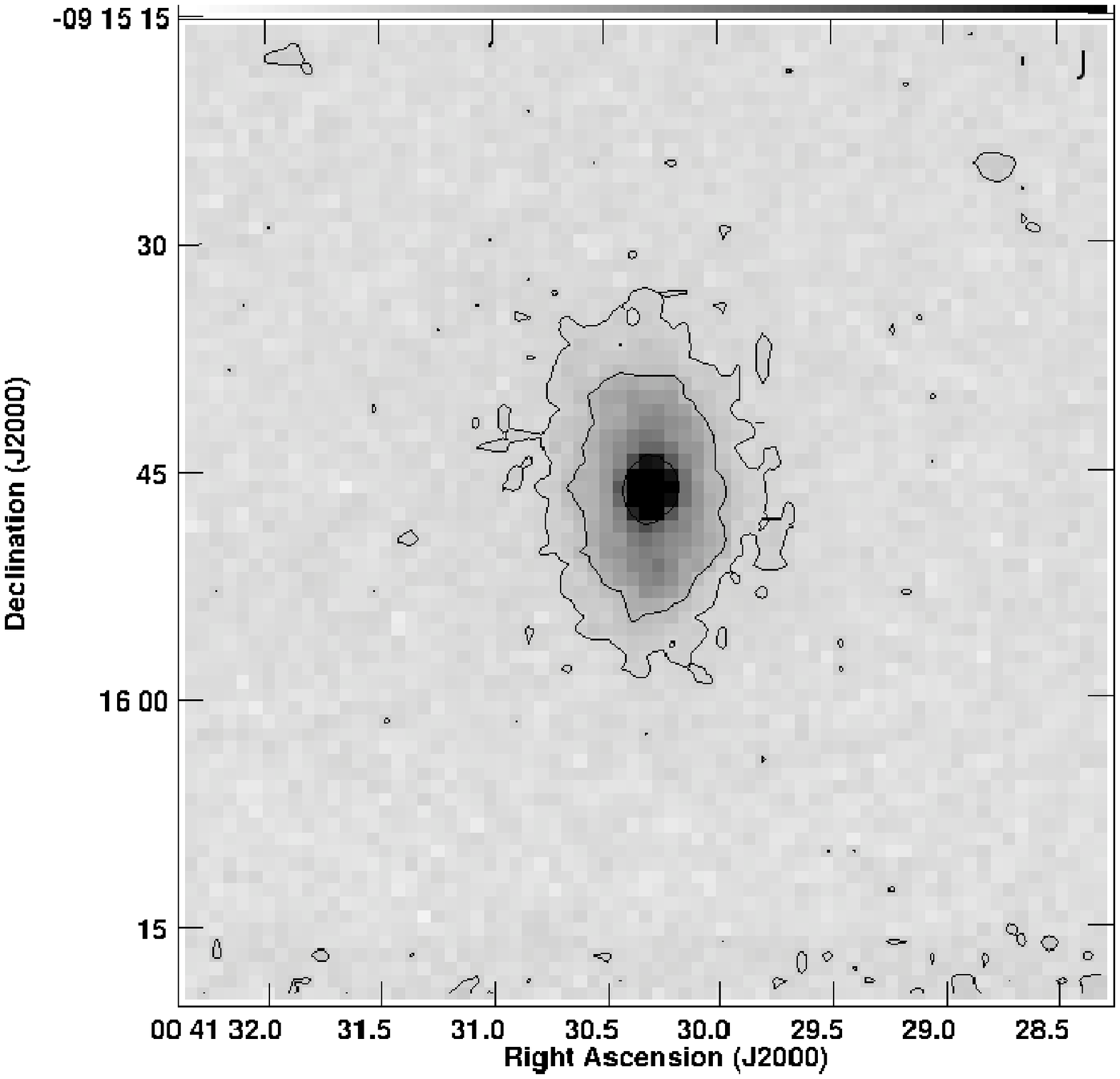}
\caption{The $jellyfish$ galaxy A85[DFL98]176 (KAZ\,364) seen in near-UV (GALEX, top left), 
optical (MEGACam-g, top right) and J-band (bottom) images.  
In spite of a slight asymmetry seen in NIR (Fig. \ref{176}), this image is not
tracing the disrupted arms extending to the SE, clearly seen in blue light.  
\label{176Jelly}}
\end{figure*}

\noindent
{\bf Field 2, a group around the jellyfish galaxy 
KAZ\,364} 
This is one of the most exciting regions of our survey.
Six bright galaxies are projected within the 3\prim\por3\prim\, 
{\sl fov} (see Fig. \ref{176}): A85[DFL98]176/186/177/174/167 
and A85[SDF98]1645.  These objects are projected onto a substructure 
named C2 (see BA09), lying \aprox8\prim\, (some 0.5\,Mpc) NW of 
the cluster center. One of these galaxies, A85[DFL98]176 
{\bf
(better known as KAZ\,364 and JO201, \cite{Bellhouse17}) 
}
is a giant spiral, 
 one of the brightest objects (in the optical bands) 
in the whole cluster. This object has a  velocity lower by
\aprox3,000\,\kms\   than the cluster systemic velocity.
Seen in blue light, this galaxy shows the pattern known
as $jellyfish$, because of the filaments of debris.
This object, with its spectacular arm disruption towards 
the east side, is included
in the $jellyfish$ sample of \citet{Pogg16}. Moreover, 
when this galaxy is seen in UV-GALEX images, clear emission 
is seen along the disrupted arms (Fig.~\ref{176Jelly}),
but the next pannels of the same figure show that the blue-distorted 
arms disappear when the galaxy is seen in the NIR.  The fact 
that the asymmetric arms are devoid of old red stars strongly 
suggests that a very strong RPS event could be at the origin 
of the stripped pattern.  The projected distance from the
cluster center (\aprox0.5\,Mpc) is well within the zone
where RPS is expected to have strong effect on gas rich
galaxies (BA09).

In addition to the disrupted arms discussed so far, on the 
east of KAZ\,364 and seen only in blue light, other 
minor asymmetries (\atf\, = 0.14) are unveiled by the NIR at the 
N and S-outskirts of the stellar disk (see Fig. \ref{176}).
These features do not seem to be linked with the disrupted 
eastern arms. No obvious neighbor could 
be blamed for a hypothetical tidal interaction:
the two closest objects in projection, A85[DFL98]177/186 
do not display strong asymmetries (0.11 and 0.07, respectively), 
and they have large radial velocities (\aprox16,328 and \aprox17,566\,\kms) 
relative to KAZ\,364 (13,393 \kms). Two other objects in the 
same field, A85[DFL98]167/174, are closer to KAZ\,364 in radial velocity
(14,167 and 13,997 \kms, respectively), and they are projected around
2\prim\, (120\,kpc), N of KAZ\,364.  In principle, a flyby interaction 
of KAZ364 with one of the galaxies seen in this field cannot be totally 
discarded.  
We conclude that both mechanisms, RPS and a minor tidal interaction,
are affecting this galaxy, producing different kinds of asymmetries.

Considering all the galaxies projected within this group, 
they seem to be part
of the loose group $C2$, where no strong tidal interactions 
seem to occur among the member galaxies, probably because 
of their high relative velocities. The slight asymmetries 
we observe could be due to gravitational interactions with 
the group and/or with the cluster potential.
\\

\begin{figure*}
\epsscale{1.9}
\plottwo{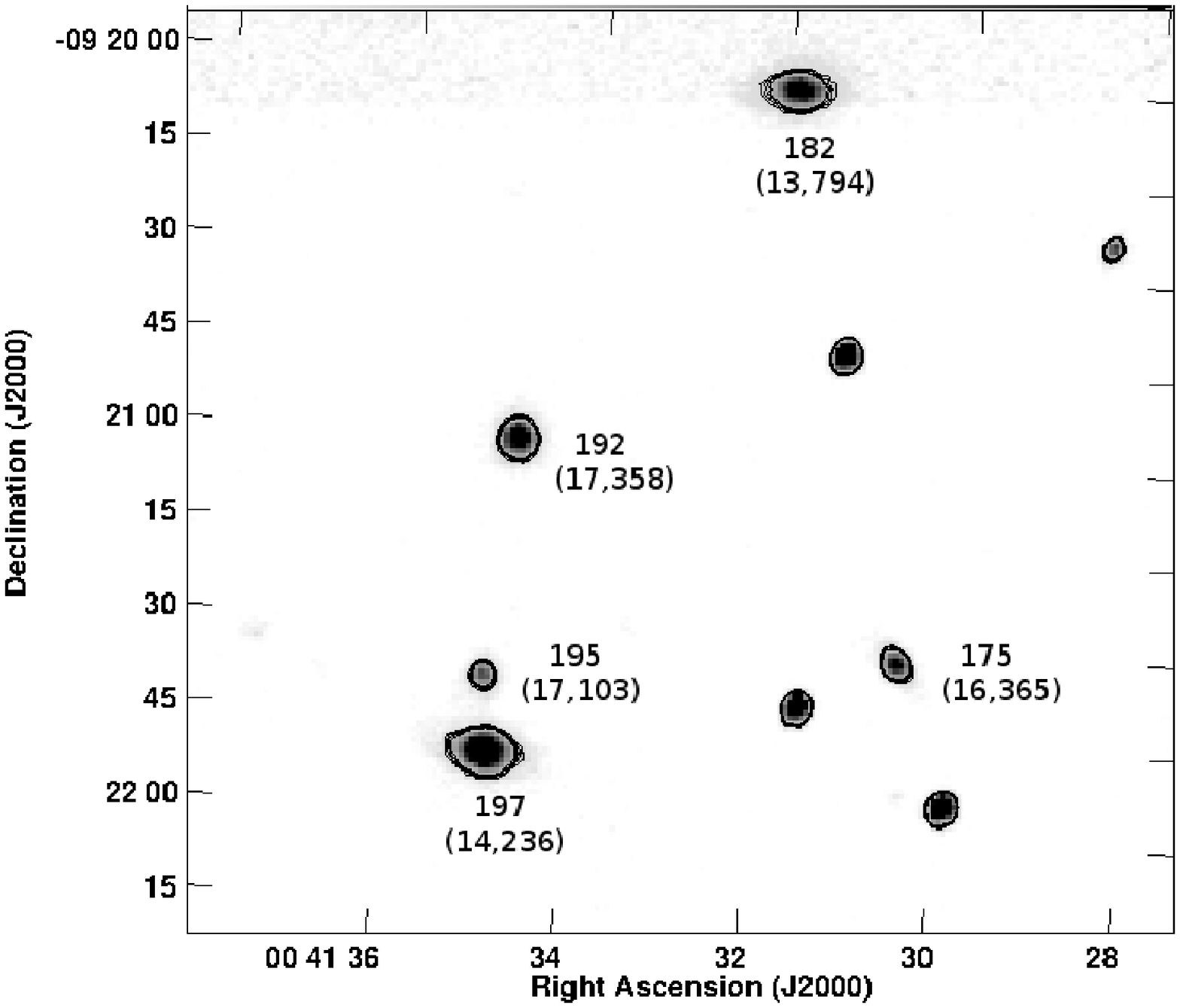}{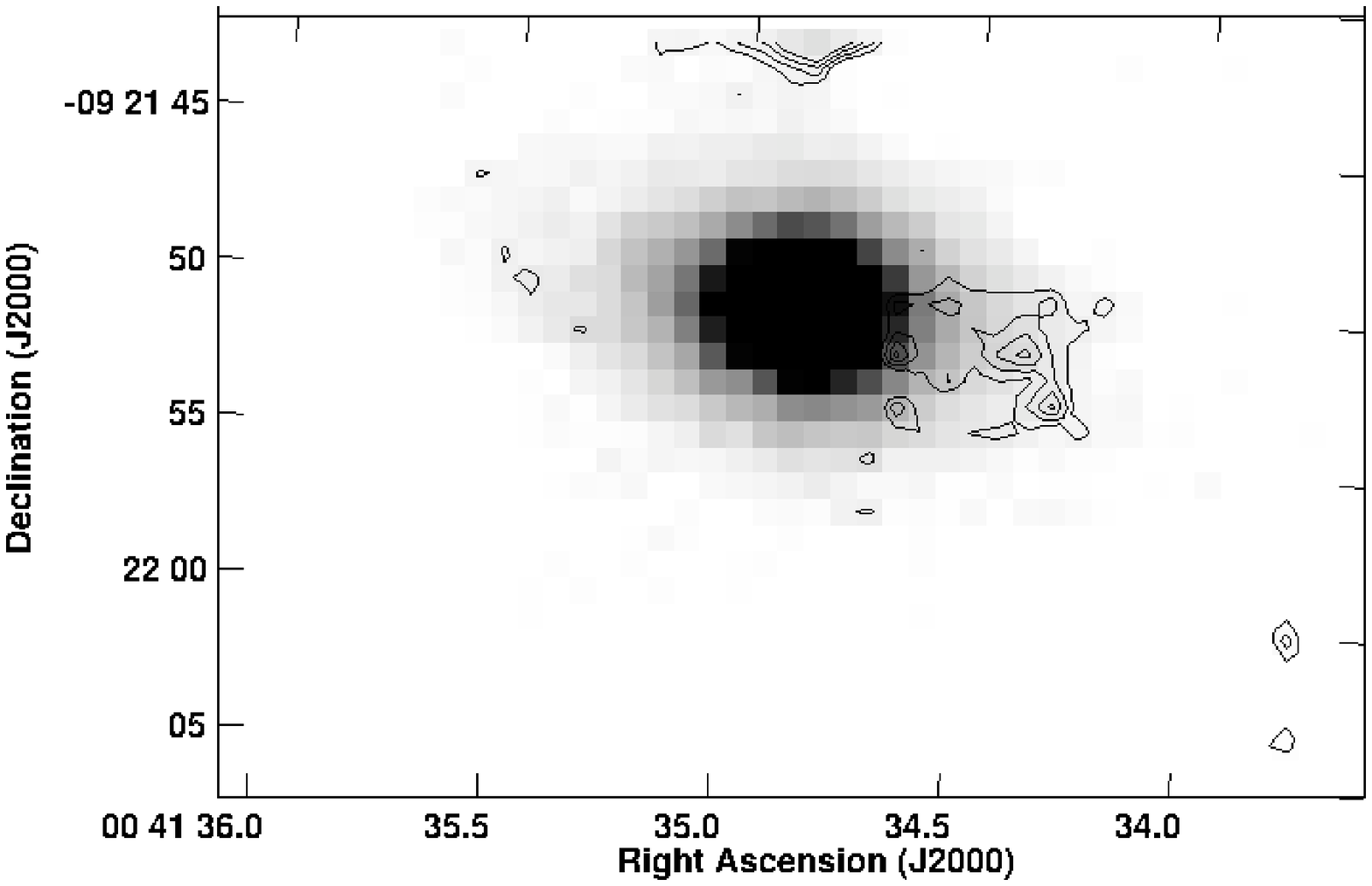}
\caption{Field 3: a $group$ of galaxies around A85[DFL98]197. Left panel: J-band in contours 
and grey scale.  The first contour corresponds to 3.5 times the $rms$ background level. 
The names of the galaxies are given according to
Table \ref{opt}.
Velocities (in \kms) are indicated in parentheses.
Right panel: close up of the galaxy A85[DFL98]197. The residual image, in contours, shows 
the asymmetries, overlaid on the J-band image (in grey scale). 
\label{197}}
\end{figure*}

\noindent
{\bf Field 3, a group around A85[DFL98]197:}   This region is
somehow similar to the previous field; five bright galaxies are 
projected close to each other, within a region  of 2\prim\,
(120\,kpc).  
The brightest object, A85[DFL98]197, displays important NIR 
asymmetries (\atf = 0.25),  strongly suggesting that this 
galaxy suffered a gravitational interaction (see Fig. \ref{197}). 
Two objects can be responsible for this.  
The first, A85[DFL98]195, is projected very close (only 0.2\prim,
or 12\,kpc), north of A85[DFL98]197, but they span a large relative
velocity of \aprox3,000\,\kms. On the other hand,  A85[DFL98]182 is
projected farther to the north (\aprox2.0\prim, 120\,kpc),
having a small difference in radial velocities (\aprox400\,\kms). 
So, a flyby interaction, some 10$^8$ yrs ago, between A85[DFL98]197
and A85[DFL98]182, could be at the origin of the observed asymmetry.
This timescale is calculated assuming the lower limit for the distance
(i.e. the projected distance between the two galaxies), and the velocity 
dispersion of A\,85 (roughly 1,000\,\kms) as a likely speed difference between 
the galaxies.
\\

\begin{figure*}
\epsscale{1.9}
\plottwo{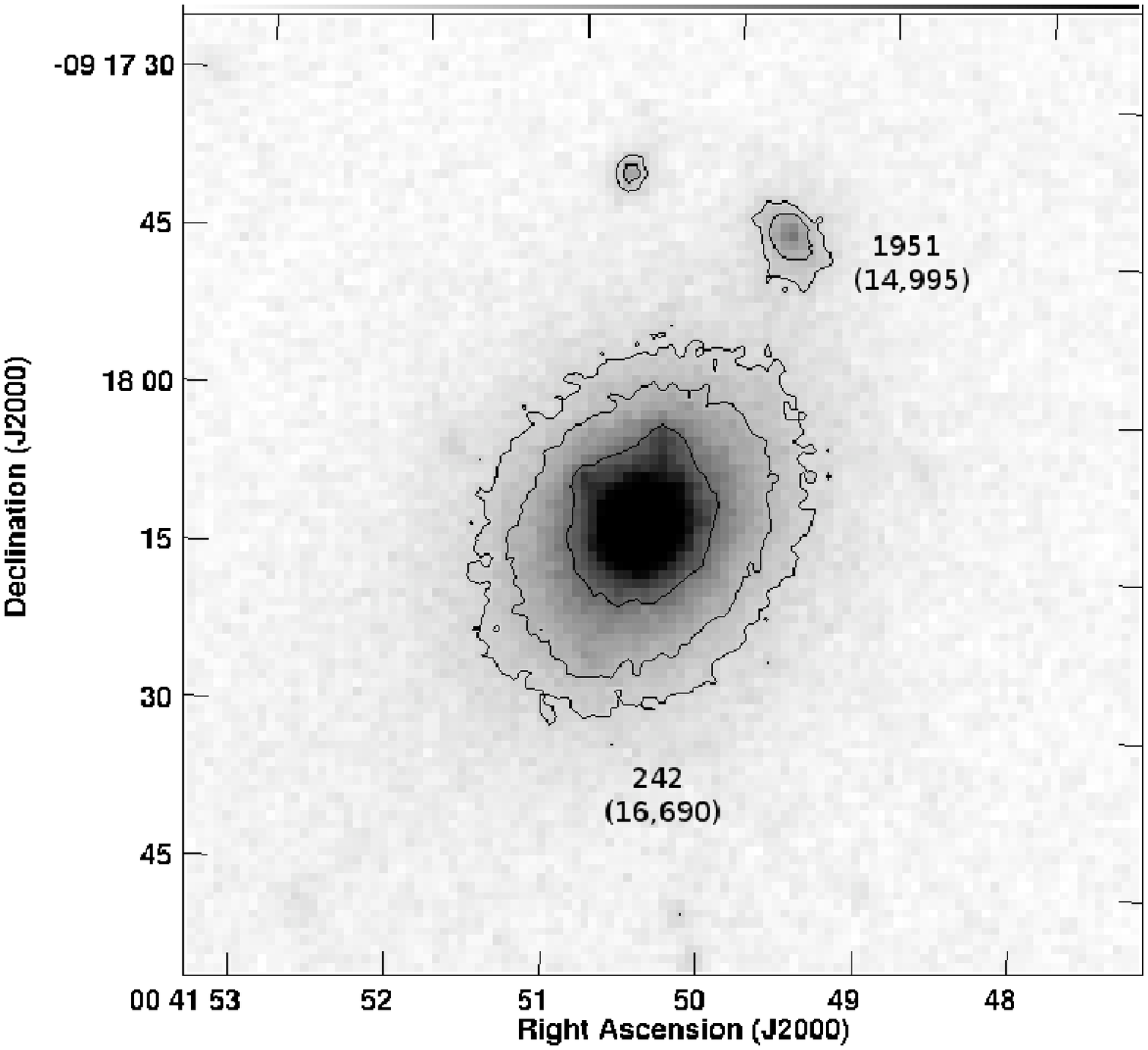}{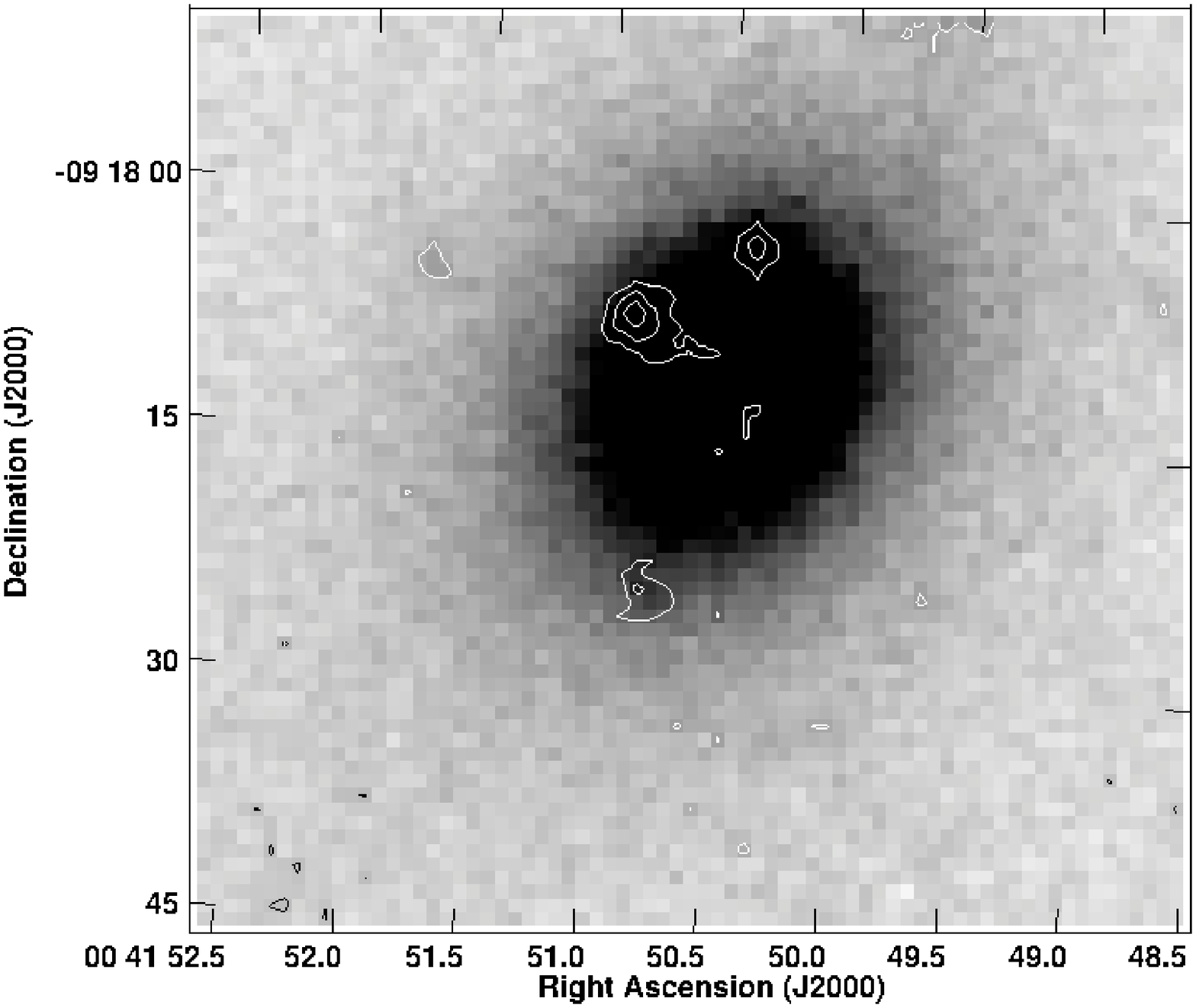}
\caption{Field 8: The cD galaxy [DFL98]242 and its satellites. Left panel: J-band in 
contours and grey scale. The first contour corresponds to 3.5 times the $rms$ 
background level.  The names of the galaxies are given according to Table \ref{opt}.
Velocities (in \kms) are given in parentheses. The low mass spiral [SDG98]1951, 
shows a degree of asymmetry through direct visual inspection.
Right panel: The residual image of the cD galaxy is shown in contours, overlaid on 
the J-band image.  This unveils three objects which are suspected to be deep in the 
cD halo. No redshifts are available for these low-mass galaxies.
\label{242}}
\end{figure*}

\noindent
{\bf Field 8, the cD galaxy A85[DFL98]242:}  This field, at the very center 
of A\,85, shows some interesting results.  While the cD appears globally 
unperturbed in the NIR, our asymmetry analysis confirmed the presence of three 
low mass galaxies, projected deep within the cD-halo, probably in the 
process of being cannibalized (see Fig. \ref{242}).  Redshifts are still to be
obtained, in order to confirm this fact. A bit farther (0.23 arcmin, \aprox14\,kpc), 
NW of the cD, the spiral A85[SDG98]1951 shows a slight asymmetry in the
NIR through visual inspection. We did not estimate the asymmetry index 
due to the small angular size. This asymmetry, seen in NIR as well as in the 
optical, suggests that this galaxy could be at an early stage of being
swallowed 
by the giant elliptical. A85[SDG98]1951 shows an abnormal red color
(see Fig. \ref{color-mag}) which could be produced by contamination by
the cD halo. 
\\

\begin{figure*}
\epsscale{0.8}
\plotone{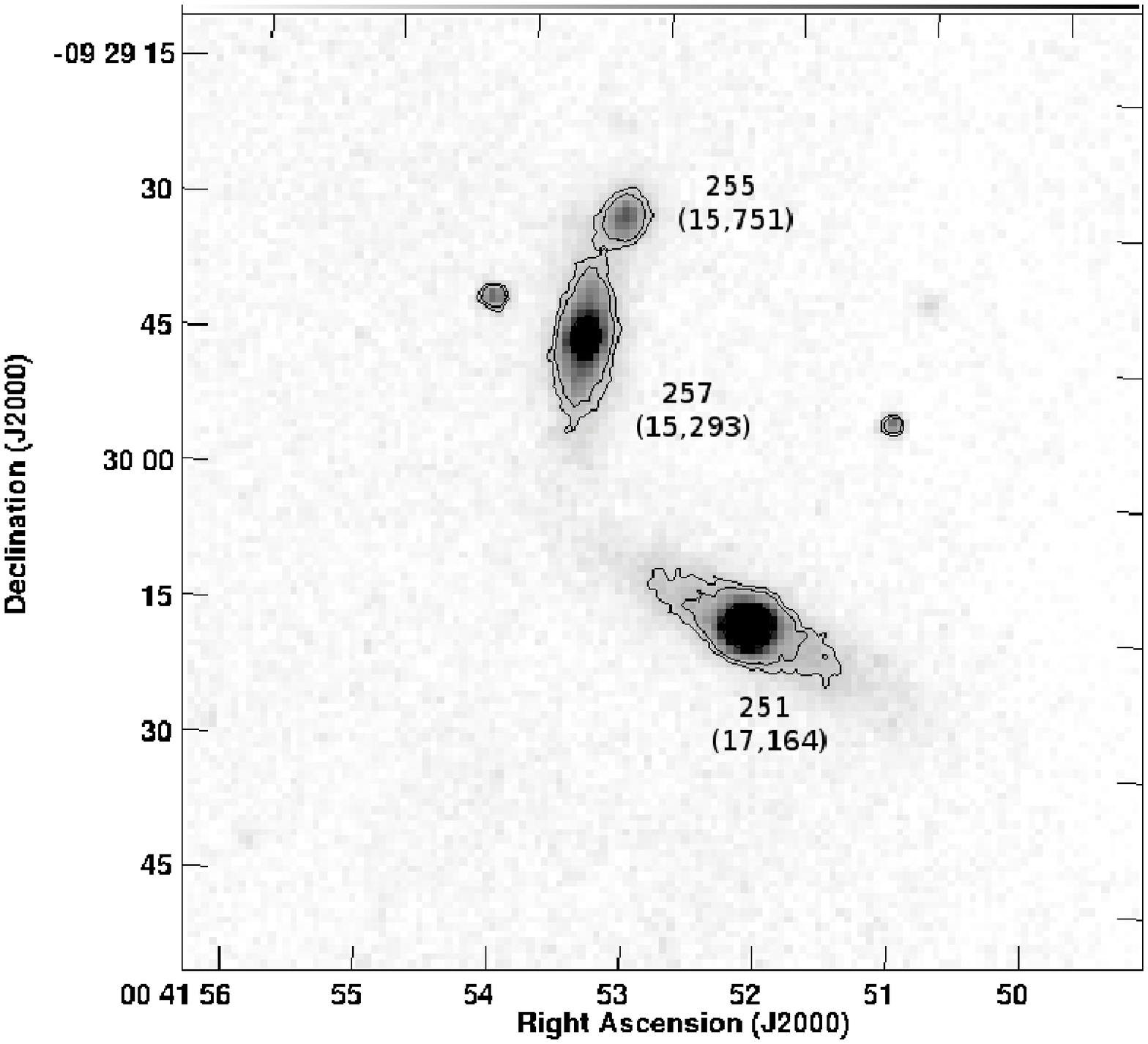}
\plotone{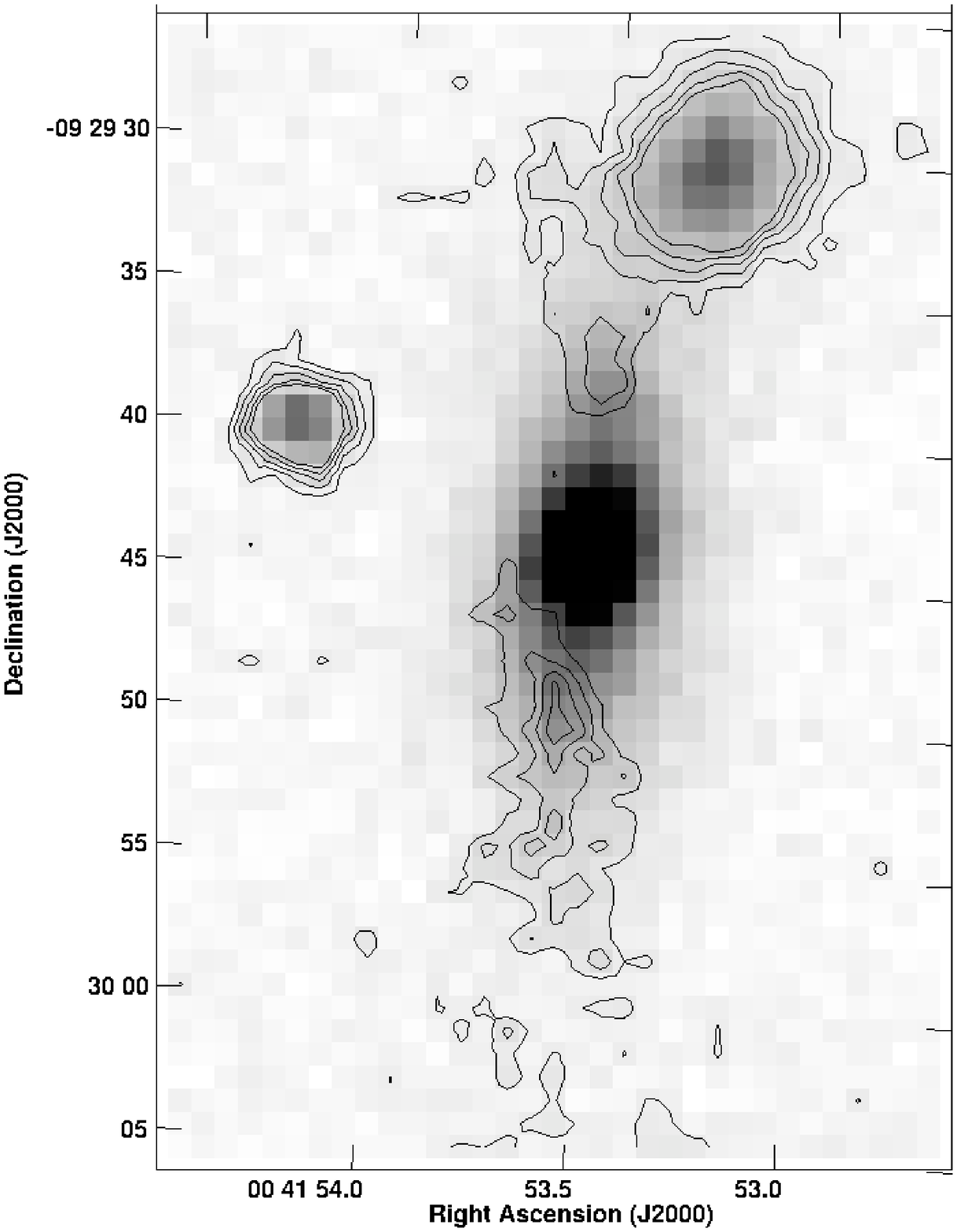}
\plotone{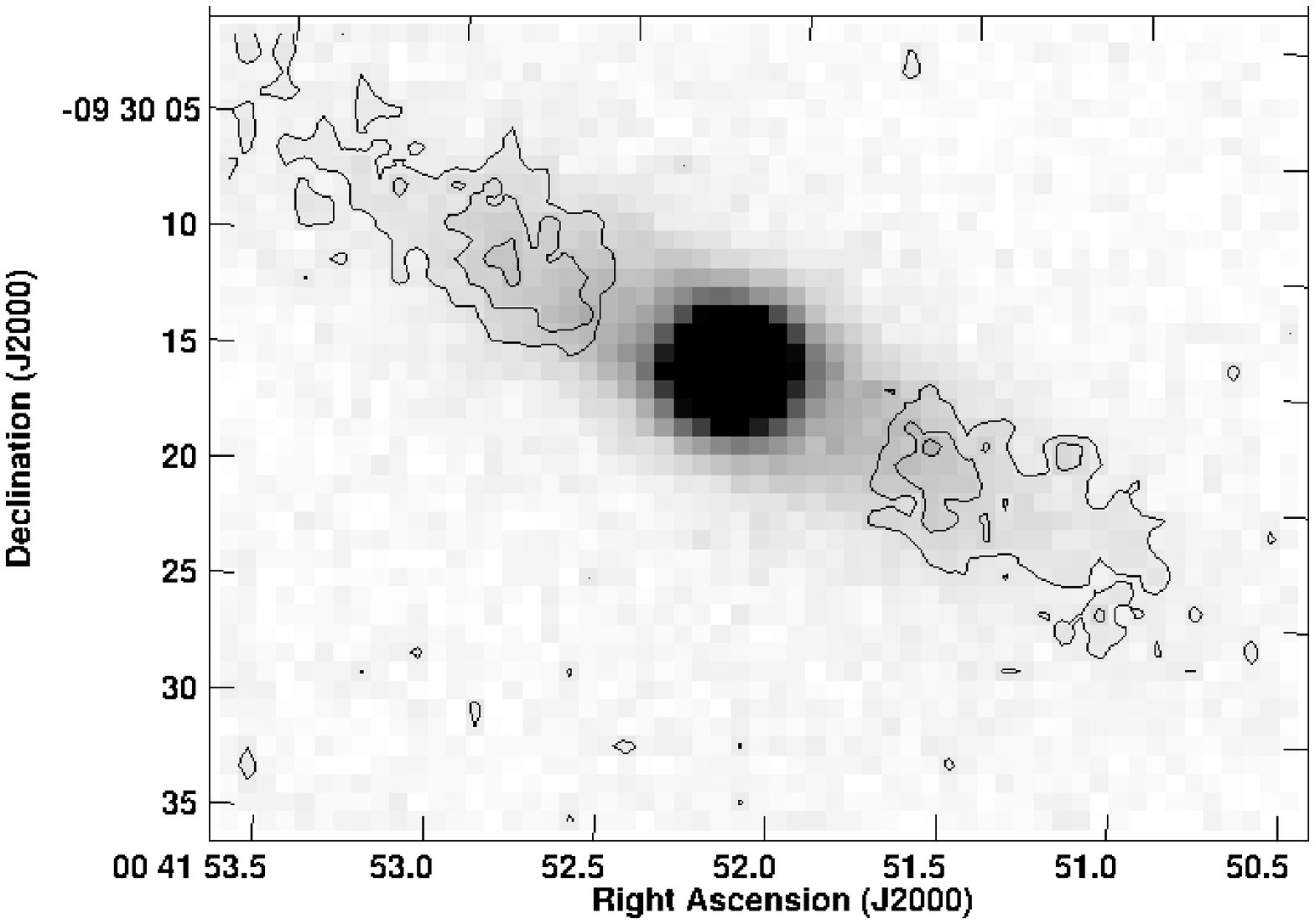}

\caption{Field 9: A triplet around [DFL98]257. Top left panel: J-band in contours 
and grey scale. The first contour corresponds to 3.5 times the $rms$ background 
level.  The names of the galaxies are given according to Table \ref{opt}.
Velocities (in \kms) are given in parentheses.
Top right: close up of A85[DFL98]257; the residual image is 
shown in contours, overlaid on the J-band image. Bottom: Residual image of
A85[DFL98]251, in contours. The strong asymmetries seen around the two massive
spirals suggests that they are part of a physical triplet, with the low mass 
galaxy A85[DFL98]255.
\label{255}}
\end{figure*}

\noindent
{\bf Field 9, the triplet A85[DFL98]251/255/257:}  This field is projected onto the
$South\,Blob$ (BA09), lying some 10\prim\, (600\,kpc) south of the cluster center.  
Three galaxies appear very close in projection from each other (see Fig. \ref{255}), 
A85[DFL98]251/257/255, two giant spirals and a low mass elliptical,
respectively.  The first one is an early spiral (Sa), lying at the SW of this 
{\it trio}; it has a radial velocity larger by 1,400\,\kms\, than the other 
two galaxies, making unclear if it is physically linked with the close pair 
A85[DFL98]255/257.  Now we show that both the large spirals (i.e. A85[DFL98]251/257)
display significant asymmetries in NIR (\atf = 0.14, 0.32, respectively), 
giving support to a recent flyby $\leq 0.25\times10^{8}$ yrs ago (estimated in 
the same way as previously).  On the other hand, 
the galaxies A85[DFL98]255/257, have a relative velocity of only \aprox450\,\kms\, 
(see Table \ref{opt}), making very likely that they constitute a 
physical pair, probably in contact. 
The southern component, the spiral A85[DFL98]257, displays enhanced 
H$\alpha$ emission (BA09), suggesting that a burst of star formation 
could have been triggered by tidal interactions with its neighbors.  
It is worth mentioning that none of the two large spirals 
in this field (A85[DFL98]251,257) were detected in \hi\, (BA09), 
down to an \hi-mass detection threshold of 7\por10$^8$\,\msolar. This
suggests that, lying well within the $South Blob$, these galaxies 
could have suffered strong RPS in addition to the observed 
tidal interactions. This could explain the absence of gas in both
spirals.
\\

\begin{figure*}
\epsscale{1.9}
\plottwo{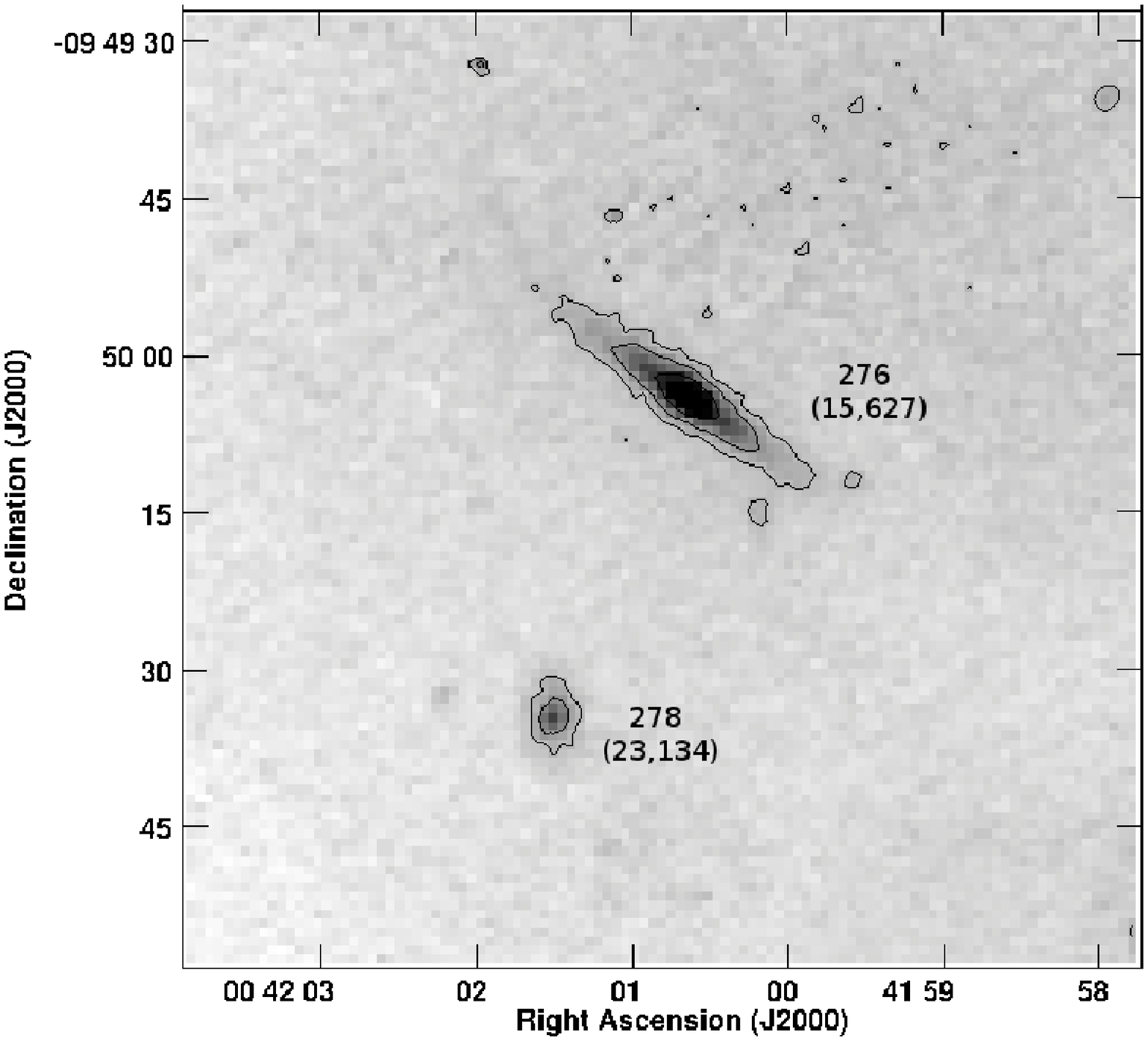}{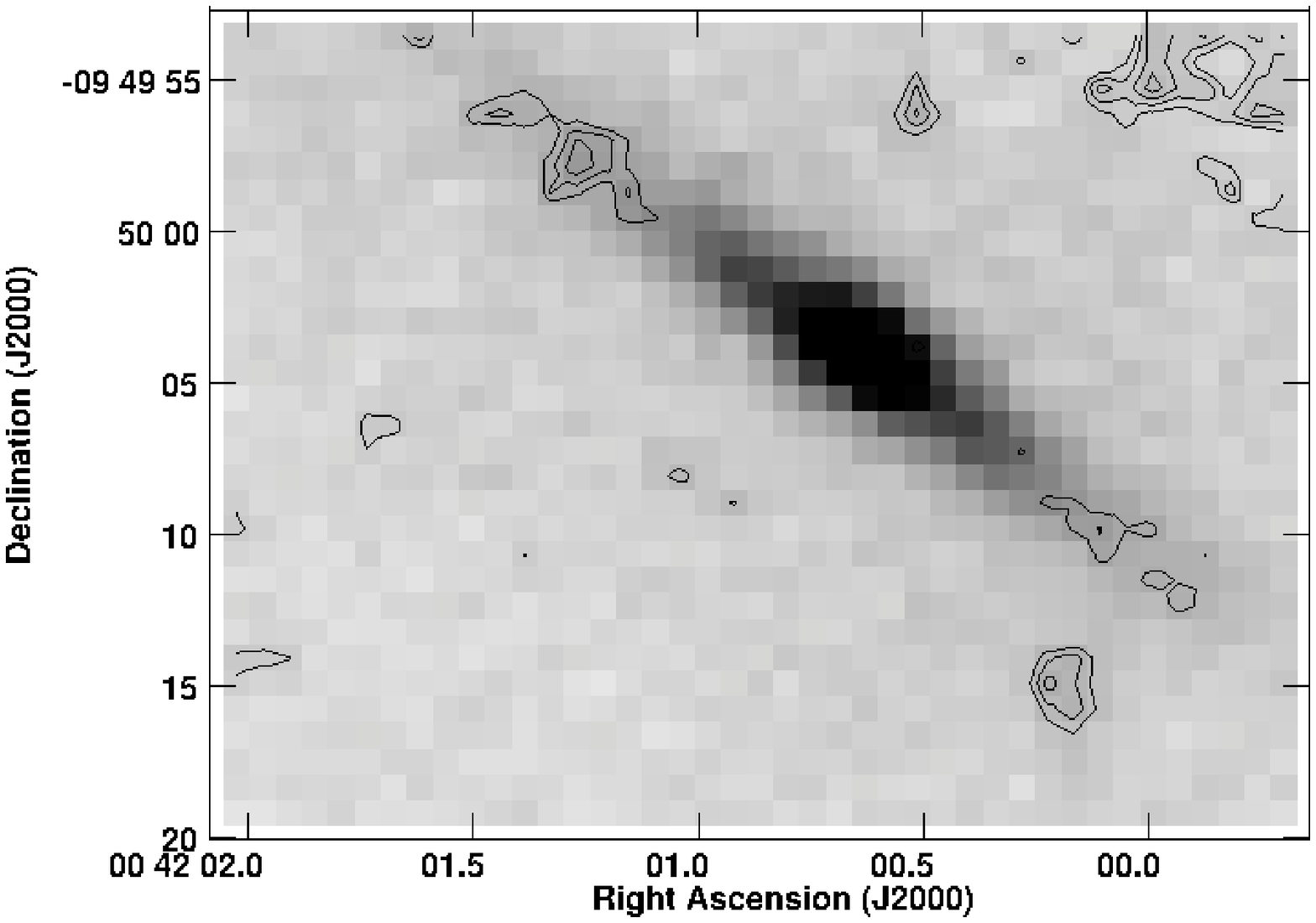}
\caption{Field 10, centered on the isolated galaxy A85[DFL98]276. 
Left panel: J-band in contours and grey scale. The first contour corresponds to 3.5 times 
the $rms$ background level.  The names of the galaxies are given according to 
Table \ref{opt}. Velocities (in \kms) are given in parentheses.
Right panel: The residual image is shown in contours, overlaid on the J-band image.
Important asymmetries are shown on both sides of the disk. 
The NW corner of the panel was affected by a slightly inhomogeneous background, 
having no effects on our asymmetry analysis.
\label{276}}
\end{figure*}

\noindent
{\bf Field 10, the isolated galaxy A85[DFL98]276:} This field hosts 
a bright (Sb) spiral. In spite of its projection, 
nearly 2\,Mpc south of the cD, and far from the detected X-ray emission, 
this galaxy is very gas deficient as no \hi\, was detected below an 
\hi-mass detection threshold of 7\por10$^8$\msolar\, (see BA09). 
Furthermore, a stellar disk with slight asymmetries on both sides, 
appears from our NIR analysis, with a larger elongation to the NE 
(see Fig. \ref{276}).  No direct neighbor can be linked to 
this galaxy, as the closest object in projection, 
(A85[DFL98]278), has a radial velocity of 23,134\,\kms. 
No cluster substructures are reported in this area, 
making this perturbed galaxy, a very intriguing one.  
This evidence suggests, assuming a radial orbit, that 
A85[DFL98]276 could be subject to galaxy harassment 
\citep{Moore96} along that cluster passage.
\\

\noindent
{\bf Fields 11-15, two very disrupted galaxies:} Several objects 
have been previously reported (BA09) in A\,85 as showing extremely 
blue colors, with only a few of them being detected in \hi. Several of 
these galaxies appear very distorted in blue light, the most striking 
cases are A85[DFL98]176 (see Field\,2, above), and A85[DFL98]286/374.

\begin{figure*}
\epsscale{0.8}
\plotone{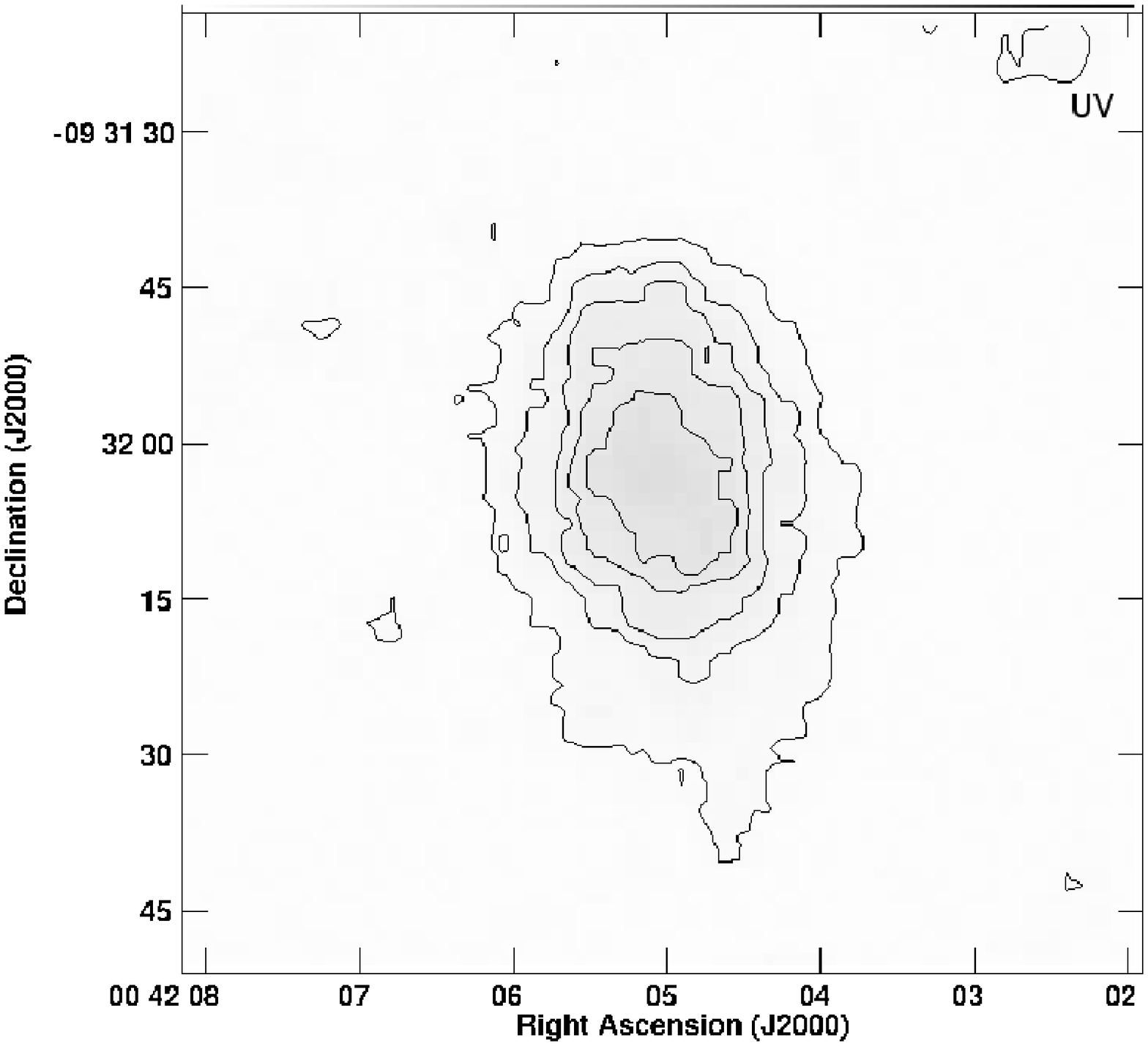}
\plotone{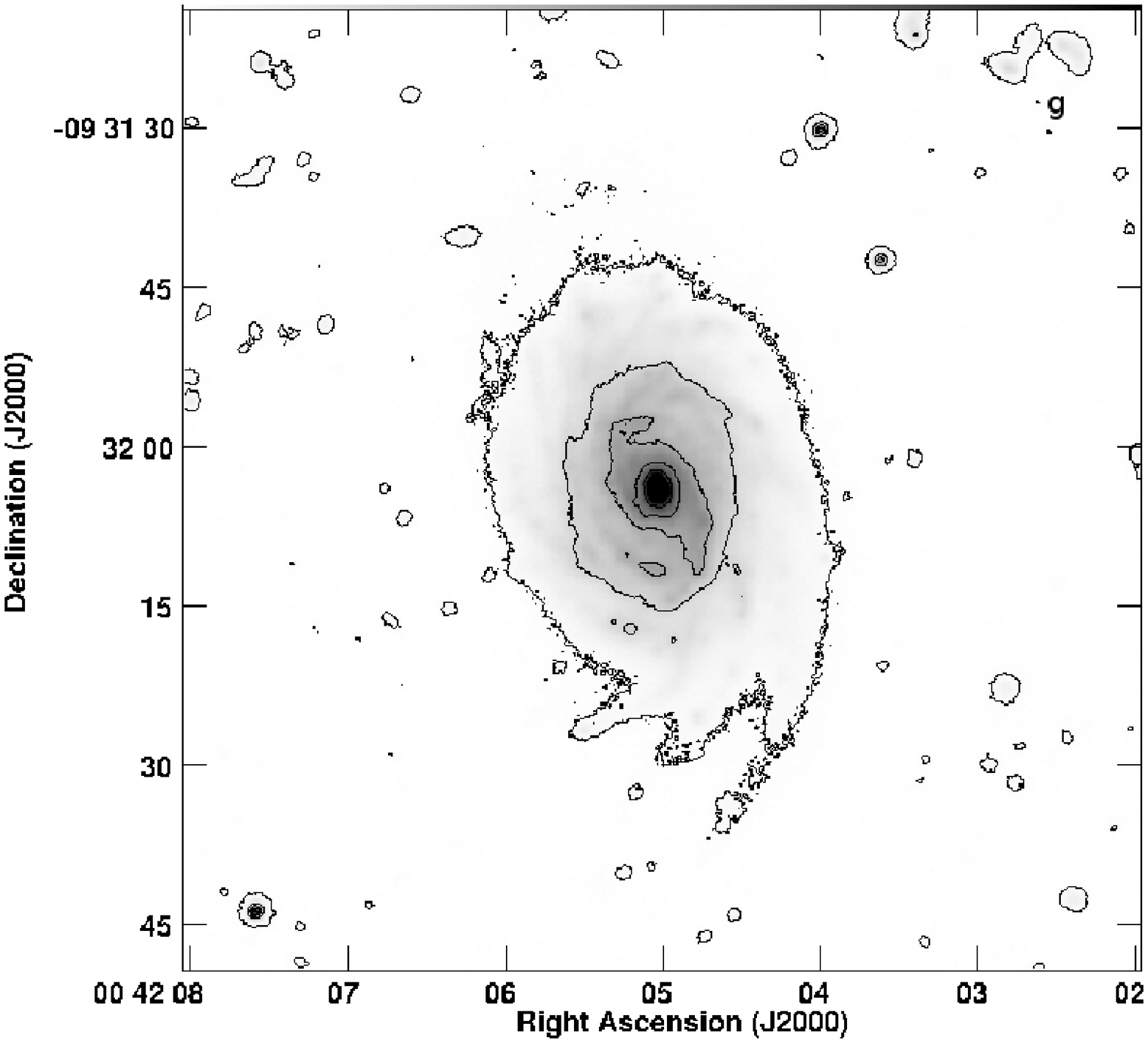}
\plotone{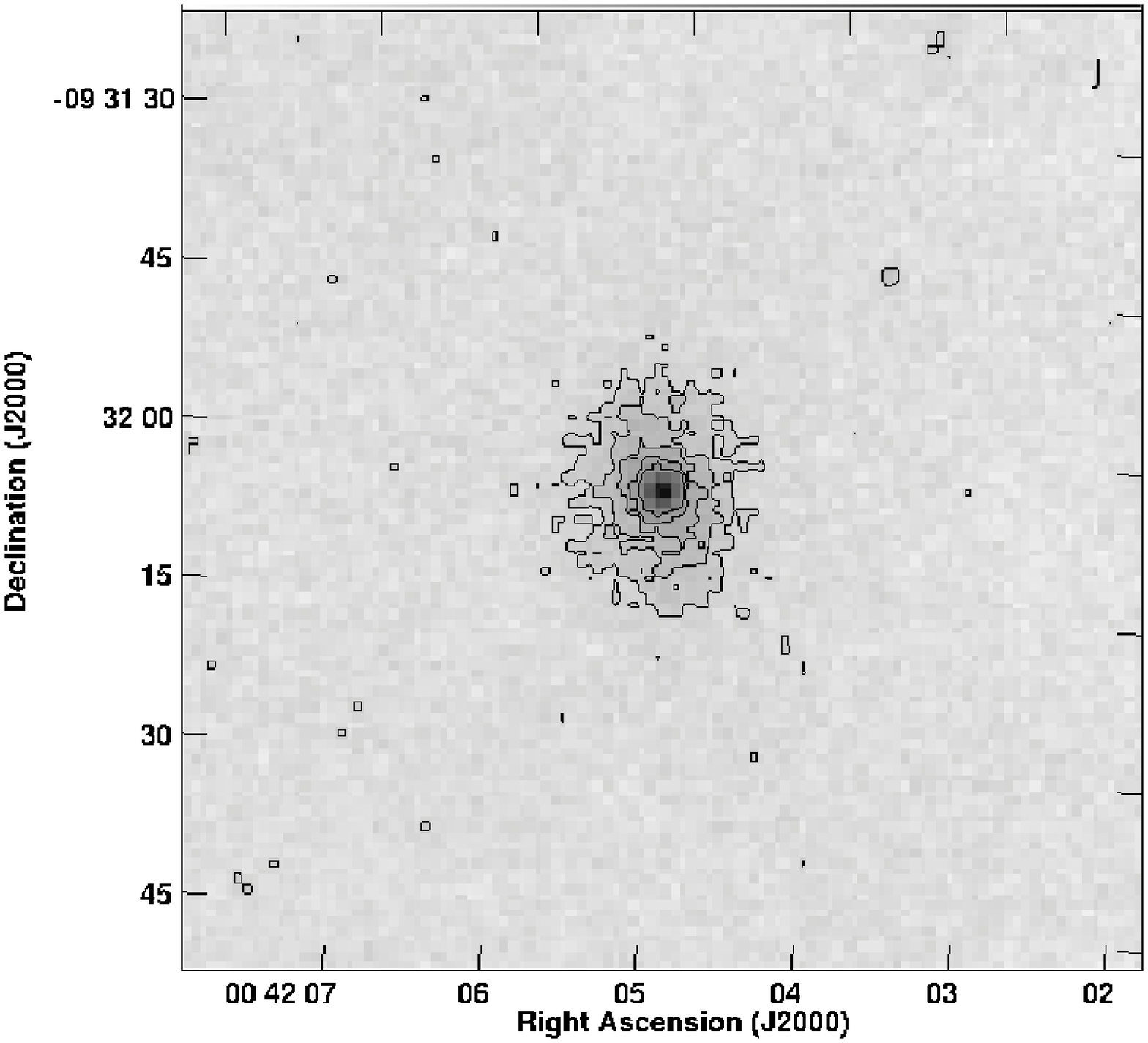}
\caption{The $jellyfish$ galaxy A85[DFL98]286 (MCG-02-02-091), seen in near-UV 
(GALEX, top left), 
optical (MEGACam-g, top right) and J-band (bottom) images.
No asymmetries are seen in the NIR, while very strongly disrupted arms 
appear in blue light to the south.
\label{286Jelly}}
\end{figure*}

A85[DFL98]286 (MCG-02-02-091) is projected onto our Field 11, 
lying 0.9\,Mpc south of the cluster center. This galaxy is a 
nearly face-on spiral, projected on the edge 
of the South Blob, within a relatively high density ICM region. 
In principle this could explain the \hi-deficiency
as it was not detected by our VLA-\hi\, survey (BA09 and 
Bravo-Alfaro et al. 2017, in prep.)
This galaxy shows disrupted arms when seen in UV and in blue images, 
and it has been cataloged as a $jellyfish$ galaxy by \citet{Pogg16}. 
Fig. \ref{286Jelly} shows that the length of the extended 
arms is shorter in A85[DFL98]286, compared with A85[DFL98]176.
Very interestingly, none of these galaxies shows old stars 
in NIR along the disrupted arms. Finally, no global asymmetry 
is obtained through our NIR analysis (\atf = 0.03); these results 
suggest that RPS is playing the most important role 
producing the strong observed disruption seen in A85[DFL98]286.

\begin{figure*}
\epsscale{0.8}
\plotone{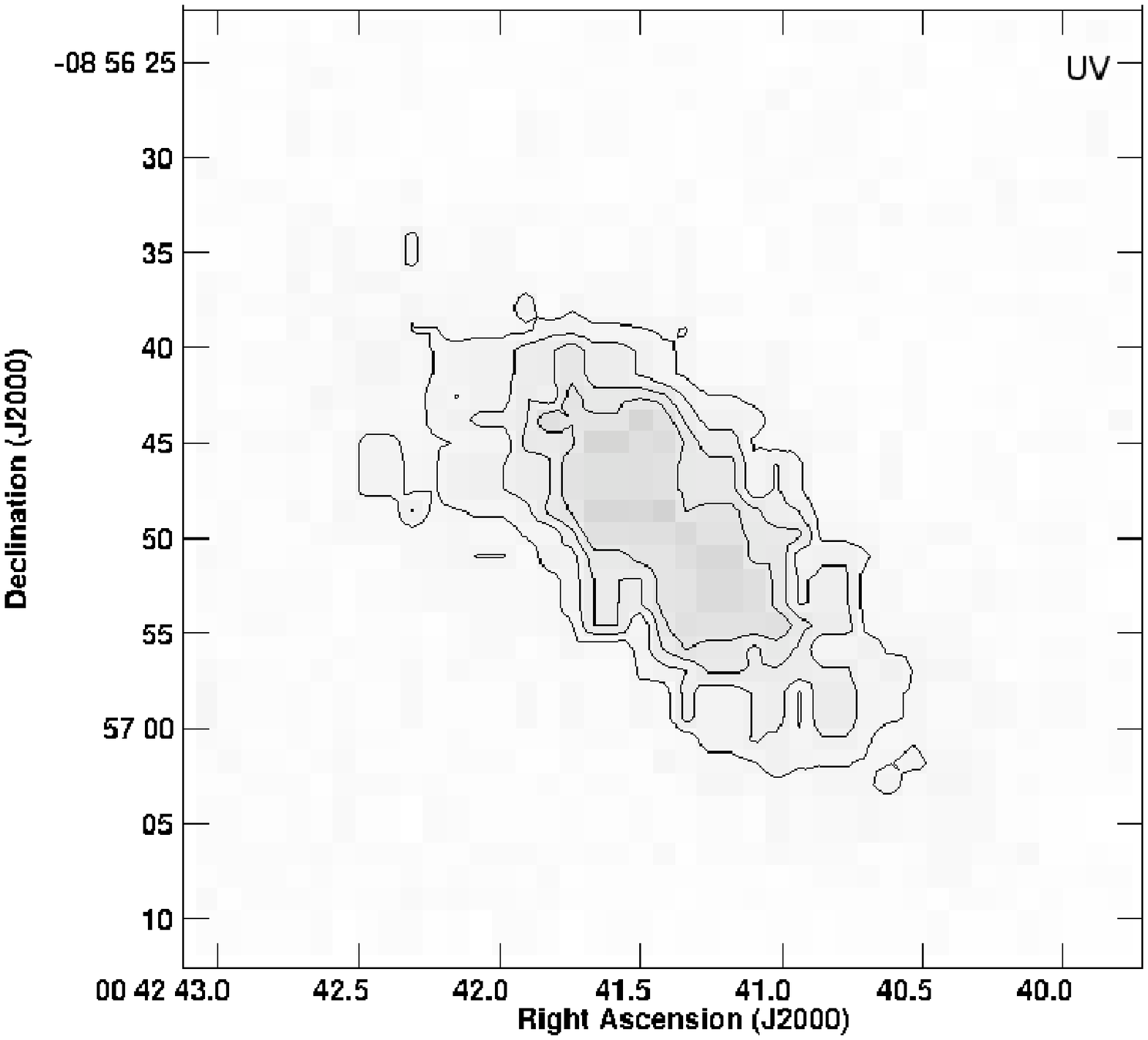}
\plotone{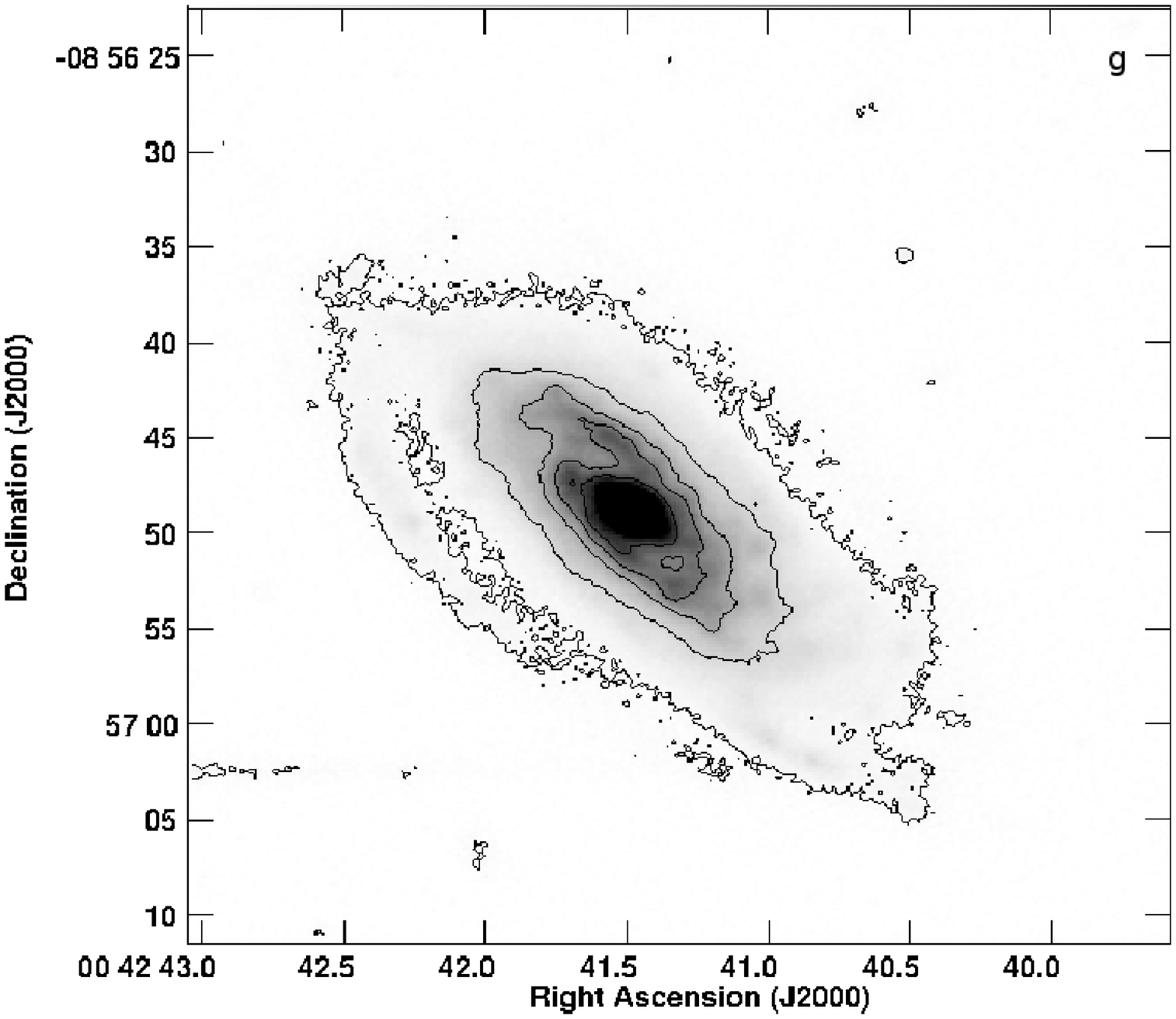}
\plotone{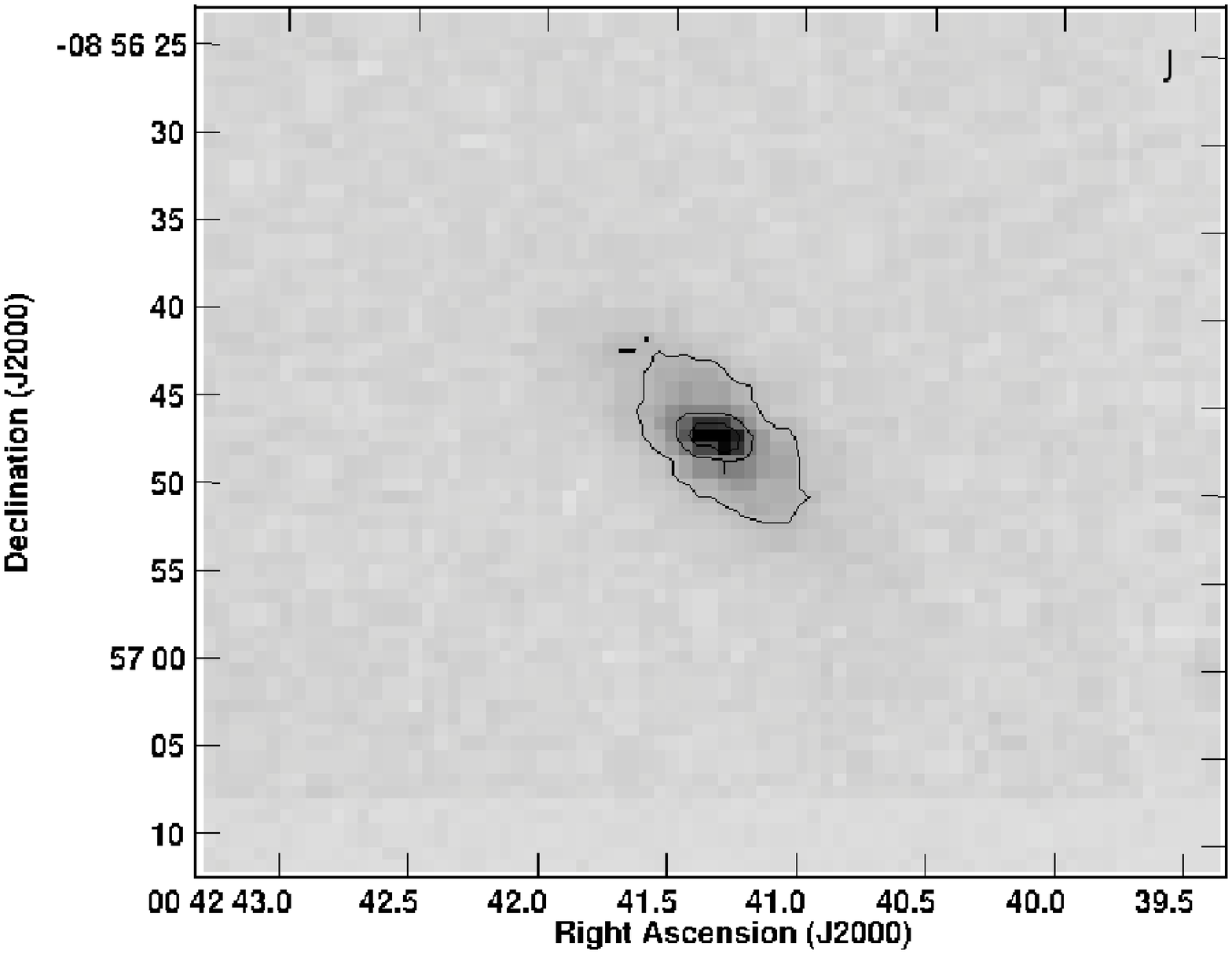}
\caption{A85[DFL98]374, a candidate to $jellyfish$, 
seen in near-UV (GALEX, top left), optical (MEGACam-g, 
top right) and J-band (bottom) images.  This is a rich \hi-galaxy,
projected 1.5\,Mpc, NE of the cluster center. A very strong variation 
of the P.A. is observed in NIR, as a function of radius. Nevertheless, 
no external asymmetries are seen in NIR compared with the 
strongly disrupted arms appearing in blue light, to the SW.
\label{374Jelly}}
\end{figure*}

Another remarkable case among the blue and disrupted galaxies is
A85[DFL98]374, which may well be a third $jellyfish$ galaxy in
A\,85.  This object was observed 
within our field 15, some 1.5\,Mpc NE of the cluster center 
(Fig. \ref{A85-fields}).  The strong disruption seen 
through visual inspection in the UV and optical bands 
(see Fig. \ref{374Jelly}), follows the pattern seen in 
the two $jellyfish$ galaxies described above.  Nevertheless, 
A85[DFL98]374 could be in an earlier stage of disruption
compared with A85[DFL98]176/286: first, the elongated 
arms in A85[DFL98]374, on the SW, are less "developed" 
and are shorter than in the other two disrupted objects. And second,
this galaxy still shows a high \hi\, content (Bravo-Alfaro 
et al. 2017, in prep.).  Concerning the NIR, A85[DFL98]374 appears
very symmetric (\atf = 0.02) and no red stars 
are seen along the disrupted arms, just like in the two 
$jellyfish$ galaxies.  So, a strong RPS event could be at
the very first stages of sweeping gas away from the disk, 
forming new stars along the gas tails.  In view of the large 
distance from the cluster center, a high speed relative to the cluster 
is needed for RPS to be efficient.   In their analysis of RPS vs 
cluster-centric distance in A85, BA09 showed that relative 
velocities above 1,000\kms\, are necessary for RPS to 
overcome the restitution force exerted on the \hi-gas,
at the projected distance of A85[DFL98]374.

\section{Summary and Conclusions} \label{concl}

{\bf 
Our main results are summarized as follows:

1. With the aim of unveiling and studying specific cases of tidally 
disrupted objects in Abell\,85, we observed 26 fields in the NIR,  
3\prim\por 3\prim\, in size, and obtained accurate J, H, K'-photometry 
for 68 bright galaxies. Our apperture NIR magnitudes are in close agreement 
with 2MASS, with our images being \aprox1\,mag\,arcsec$^{-2}$
deeper. Our J, H, K' atlas of images are available upon request.
}

2. With the aim of providing quantitative information on the
presence (and degree) of tidal disruptions,  we propose
a new asymmetry index, \atn.  From the sample of 68 galaxies,
we selected the 41 largest in angular size, in order to go 
through an asymmetry analysis.  Our index is able to measure
(in surface) the asymmetry features in a galaxy.
This tool proved to deliver important complementary 
information to that provided by other indices available in the 
literature.  

3. Among 41 bright galaxies going through our asymmetry analysis
we report 10 objects showing mild-to-strong asymmetries.
For a few of the disrupted objects, asymmetries could be 
seen through visual inspection on our NIR images. Nevertheless,
our method unveiled unexpected asymmetry features associated
with other galaxies, confirming the efficiency of the
residual technique.  We quantified the degree of asymmetry
with the \atf\, index, finding that these perturbations go 
from mild (\atf\,=\,1.0) to strong (1.1$\leq$\,\atf\,$\leq$\,0.32). 
We compared the residuals coming from the $bmodel$ and the
180\degree-rotation method, and found that the first method delivers a
systematically higher asymmetry index.
Even considering our biased sample, it is important to notice
that the fraction of disrupted galaxies among the brightest objects
of A\,85, is already close to 25\%. This confirms that gravitational 
mechanisms are playing a role in transforming galaxies in this cluster.

4. We combined our NIR study with previous results of substructures 
found in A\,85. The asymmetries measured in the NIR allowed to confirm 
the presence of some physical pairs and groups, linked with larger
structures. For instance, galaxies observed in our Fields 2 and 3, 
are projected onto the same substructure, $C2$ (BA09), some 200-300 kpc
west of the cluster center. If we consider that this structure is
believed to be infalling from the background with a high velocity 
relative to the cluster, then the galaxies within this group would 
be undergoing galaxy $pre-processing$ before reaching the main 
cluster body, accounting for the slight asymmetries observed in NIR.
Since the  velocity dispersion among the objects within
this group is large (above 1,000 \kms), they might constitute a loose 
group of galaxies. Another case is observed within our Field 9,
where three galaxies are projected within the $South\,Blob$ (BA09).  
The significant NIR asymmetries, measured on the two giant spirals, 
strongly suggest that they have been in contact, probably through a
flyby interaction, less than 10$^8$\,years ago.

5. A very interesting issue we approached in this paper was the 
deep NIR imaging of three very disrupted (two of them
being classified as $jellyfish$) galaxies
in A\,85: A85[DFL98]176/ 286/374.  We have shown that comparing 
the NIR morphology with the UV-optical delivers very useful 
physical information about such disrupted galaxies.  The three 
objects display different degrees of morphological disruption,
A85[DFL98]176 being the most dramatic case.  This kind of galaxies
are well known to display disrupted arms, being very 
bright in UV and blue bands. 
We have shown that the disrupted arms are not detected 
in the NIR bands, in spite of our deep images going down to 
22.4\,mag\,arcsec$^{-2}$ (in J-band). 
This absence of old stars along the disrupted 
arms discards any tidal interaction as the origin of the 
perturbation: gravitational interactions would tear up all 
kind of stars from the galaxy disk, both blue and red ones. 
Our results support the hypothesis that a very strong RPS event,
observed at different stages along the three objects, is 
responsible for the galaxy disruption and formation of the 
arms/tails. In this scenario
RPS removed a large fraction of the \hi-gas, and 
the bright stars seen in UV-optical are formed along
the gas tails.  \\

We have shown that combining deep NIR imaging with other
datasets, such as optical imaging and redshifts, as well
as substructures in clusters, constitutes a powerful tool
to investigate the recent evolution of galaxies infalling 
into such massive systems. We have also shown that measuring 
asymmetries allows to quantify the degree of interaction
a galaxy is undergoing. All this sheds light on the role 
played by environment, and by different physical mechanisms
driving the infall and evolution of galaxies in clusters. In our 
forthcoming papers we will combine detailed \hi\, information
(maps, gas content, kinematics) with homogeneous optical/NIR
imaging, both covering large volumes of a sample of nearby clusters.

\vspace{2cm}

\begin{acknowledgments}
We thank the anonymous referee for her/his suggestions which helped
to improve this paper.  HBA acknowledges CONACyT grant 169225, 
and the Institute d'Astrophysique de Paris, for the invitation to 
carry out a one-year working stay.   This work was supported by 
Funda\c{c}\~ao para a Ci\^encia e a Tecnologia (FCT, Portugal) through 
national funds (UID/FIS/04434/2013) and by FEDER through COMPETE2020 
(POCI-01-0145-FEDER-007672).  F.D. acknowledges long-term support 
from CNES. MLG and YV acknowledge the support provided by
DAIP, at UGTO. This research has made use of the NASA/IPAC 
Extragalactic Database (NED) which is operated 
by the Jet Propulsion Laboratory, California 
Institute of Technology, under contract with 
the National Aeronautics and Space Administration. 

\end{acknowledgments}

\acknowledgments

\begin{deluxetable}{ccccrcl}
\tabletypesize{\scriptsize}
\tablecaption{The fields observed in NIR in A\,85\label{fld}}
\tablewidth{0pt}
\tablehead{
\colhead{Field} & \colhead{$\alpha_{J2000}$} & \colhead{$\delta_{J2000}$} & \colhead{Year} 
& \colhead{ Obj } & \colhead{$t$ (sec)} & {Notes}\\
\colhead{} & \colhead{} & \colhead{} & \colhead{} & \colhead{} & \colhead{J, H, K'}   \\
\colhead{(1)} & \colhead{(2)} & \colhead{(3)} & \colhead{(4)} 
& \colhead{(5)} & \colhead{(6)}  & \colhead{(7)}
}
\startdata
{1} & {00 41 19.8} & { -09 23 27} & {2007, 2009} & {150} & 3260 3240 3380 & pair/\hi-def\\
{2} & {00 41 30.3} & { -09 15 46} & {2007, 2009} & {176} & 3480 3240 3510 & group\\
{3} & {00 41 35.1} & { -09 21 52} & {2009} & {197} & 2280 2340 3780 & group\\
{4} & {00 41 36.1} & { -08 59 36} & {2010,2011} & {201} & 2700 2700 3240 & pair/\hi-def \\
{5} & {00 41 39.6} & { -09 14 57} & {2010} & {209} & 2160 2160 2700 & pair \\
{6} & {00 41 40.1} & { -09 18 15} & {2009} & {210} & 1620 1620 3780 & pair\\
{7} & {00 41 43.0} & { -09 26 22} & {2009} & {221} & 1800 1980 2640 & group/\hi-def\\
{8} & {00 41 50.5} & { -09 18 11} & {2009} & {242} & 1620 1740 2640 & cD\\
{9} & {00 41 53.2} & { -09 29 29} & {2006,2011} & {255}& 2520 2520 2520 & group/\hi-def\\
{10} & {00 42 00.6} & {-09 50 04} & {2006} & { 276} & 2640 2640 3600 &  isolated\\
{11} & {00 42 05.0} & { -09 32 04} & {2006,2011} & {286} & 3840 3840 2850 & pair/\hi-def \\
{12} & {00 42 18.7} & {-09 54 14} & {2006,2010} & {323} & 2700 3240 3240 & \hi-rich\\
{13} & {00 42 24.2} & { -09 16 17} & {2010} & { 338} & 2160 2160 2700 & pair/\hi-def\\
{14} & {00 42 29.5} & { -10 01 07} & {2006} & { 347} & 3600 3975 3540 & \hi-rich\\
{15} & {00 42 41.5} & { -08 56 49} & {2007} & { 374} & 2630 3490 3655 & group/\hi-rich\\
{16} & {00 42 43.9} & { -09 44 21} & {2011} & { 382} & 2160 2160 2160 & pair/\hi-def\\
{17} & {00 42 48.4} & { -09 34 41} & {2011} & { 391} & 2160 2160 2160 & \hi-def\\
{18} & {00 43 01.6} & { -09 47 34} & {2006,2010} & {426}& 3804 3480 3240 & group/\hi-rich\\
{19} & {00 43 10.1} & { -09 51 41} & {2006,2011} & { 442}& 3000 3800 2890 & group/\hi-def\\
{20} & {00 43 11.6} & { -09 38 16} & {2006} & { 451} & 3300 3000 3000 & pair/\hi-def\\
{21} & {00 43 14.3} & { -09 10 21} & {2007} & { 461} & 2430 4100 3700 & blue/\hi-rich\\
{22} & {00 43 19.5} & { -09 09 13} & {2007} & {*3114}& 3600 3600 3600 & blue/\hi-rich\\
{23} & {00 43 31.2} & { -09 51 48} & {2006} & { 486} & 3800 3800 2840 & blue/\hi-rich\\
{24} & {00 43 34.0} & { -08 50 37} & {2007} & {491} & 3240 3800 3800 & blue/\hi-rich\\
{25} & {00 43 38.7} & { -09 31 21} & {2006} & { 496}& 3780 3660 3720 & blue/\hi-rich \\
{26} & {00 43 43.9} & { -09 04 23} & {2007} & {502}& 3240 3500 3600  & blue/\hi-rich\\
\enddata
\tablecomments{Column (1): the field number, ordered by R.A.
Columns (2) and (3): the center of each field. Column (4): the year(s) 
of the corresponding observing run. Column (5): the galaxy used as reference 
within each field; names are taken from 
\citep{Durret}, except (*), coming from \citep{Slezak}. 
Column (6): total integration times, for each band, in seconds. 
Column (7): Notes about the interest associated with each field; see text.}
\end{deluxetable}

\begin{deluxetable}{cccclcc}
\tabletypesize{\scriptsize}
\tablecaption{Optical data of the observed galaxies in A\,85\label{opt}}
\tablewidth{0pt}
\tablehead{
\colhead{Field} & \colhead{Galaxy} &\colhead{$\alpha_{2000}$, $\delta_{2000}$} & \colhead{Vel.} 
&\colhead{Opt.}& \colhead{Diam.} & \colhead{Morph.}\\
\colhead{} & \colhead{} &\colhead{} & \colhead{(km/s)} 
&\colhead{magn.}& \colhead{(\prim)} & \colhead{}\\

\colhead{(1)} & \colhead{(2)} & \colhead{(3)} & \colhead{(4)} 
& \colhead{(5)} & \colhead{(6)} & \colhead{(7)}
}
\startdata
{1} & {145} & {00 41 19.0, -09 23 24} & {14,935} & {17.9} & {0.21}   & {-}\\
    & {150} & {00 41 19.8, -09 23 27} & {14,681} &  {16.5r} & {0.30}  & {-}\\
\hline
{2} & {167} & {00 41 27.1, -09 13 42} & {14,167} & {16.7} & {0.25}   &  {-}\\
    & {*1645} & {00 41 27.9, -09 13 47} & {16,315} & {17.1} & {0.78} &  {-}\\
    & {174} & {00 41 28.8, -09 13 59} & {13,997} & {15.4v} & {0.50}  &  {-}\\
    & {176} & {00 41 30.3, -09 15 46} & {13,393} & {15.1} & {0.37}   &  {cD*}\\  
    & {177} & {00 41 30.4, -09 14 07} & {16,328} & {15.5v} & {0.20}  &  {-}\\
    & {186} & {00 41 33.3, -09 14 57} & {17,566} & {16.9} & {0.36}   &  {-}\\
\hline
{3} & {175} & {00 41 30.5, -09 21 33} & {16,365} & {17.7} & {0.34}   &  {-}\\
    & {182} & {00 41 32.0, -09 20 03} & {13,794} & {16.3} & {0.37}  &  {-}\\
    & {192} & {00 41 34.7, -09 21 00} & {17,358} & {16.3} & {0.24}   &  {-}\\
    & {195} & {00 41 34.9, -09 21 38} & {17,103} & {18.1} & {0.19}  &  {-}\\
    & {197} & {00 41 35.0, -09 21 51} & {14,236} & {16.6} & {0.46}  &  {-}\\
\hline
{4} & {193} & {00 41 34.9, -09 00 47} & {17,556} & {17.4} & {0.26}   &  {-}\\
    & {201} & {00 41 36.2, -08 59 35} & {17,935} & {16.8} & {0.56}  &  {-}\\
\hline
{5} & {206} & {00 41 39.0, -09 27 48} & {17,126} & {18.4} & {0.19}   &  {-}\\
    & {209} & {00 41 39.6, -09 27 31} & {16,666} & {17.6} & {0.30}   &  {-}\\
\hline
{6} & {202} & {00 41 36.2, -09 19 30} & {16,371} & {17.3} & {0.21}    &  {-}\\
    & {210} & {00 41 40.1, -09 18 15} & {16,825} & {17.5} & {0.20}   &  {-}\\
    & {214} & {00 41 41.3, -09 18 57} & {14,283} & {16.5} & {0.40}   &  {-}\\
\hline
{7} & {215} & {00 41 41.4, -09 26 21} & {16,305} & {18.4} & {0.14}   &  {-}\\
    & {221} & {00 41 43.0, -09 26 22} & {16,886} & {14.8} & {1.00}   &  {-}\\
    & {222} & {00 41 43.5, -09 25 30} & {16,923} & {18.3} & {0.21}   &  {-}\\
    & {243} & {00 41 50.2, -09 25 47} & {17,349} & {15.8} & {0.53}   &  {E}\\
\hline
{8} & {{*}1895} & {00 41 45.5, -09 16 35} & {\nodata} & {19.8} & {-}     &  {-}\\
    & {236} & {00 41 48.2, -09 17 03} & {15,870} & {16.3} & {0.35}   &  {-}\\
    & {{*}1951} & {00 41 49.6, -09 17 43} & {14,995} & {16.0} & {0.21}&  {-}\\
    & {242} & {00 41 50.5, -09 18 11} & {16,690} & {14.7b} & {1.30}   &  {cD}\\
    & {{*}1966} & {00 41 50.7, -09 17 39} & {16,536} & {18.8v} & {-}&  {-}\\
\hline    
{9} & {238} & {00 41 49.1, -09 29 03} & {18,367} & {17.0r} & {0.18}   &  {-}\\
    & {251} & {00 41 52.1, -09 30 15} & {17,164} & {14.5r} & {0.30}   & {Sa}\\
    & {254} & {00 41 53.1, -09 31 16} & {17,121} & {17.6i} & {0.24}   &  {-}\\
    & {255} & {00 41 53.2, -09 29 29} & {15,751} & {16.2v} & {0.47}   &  {E}\\
    & {257} & {00 41 53.5, -09 29 44} & {15,293} & {16.0} & {0.72}   & {Sc}\\
\hline    
{10}& {276} & {00 42 00.6, -09 50 04} & {15,627} & {16.4} & {0.81}   &  {Sb}\\
    & {278} & {00 42 01.5, -09 50 35} & {23,134} & {17.5} & {0.26}   &  {S}\\
\hline   
{11}& {286} & {00 42 05.0, -09 32 04} & {15,852} & {15.9} & {0.68}   &  {Sc}\\
    & {{*}2260} & {00 42 08.3,-09 31 05} & {16,963} & {17.8r} & {0.19}&  {-}\\
\hline    
{12}& {315} & {00 42 16.1, -09 54 28} & {38,609} & {18.3} & {0.18}   &  {S0}\\
    & {323} & {00 42 18.7, -09 54 14} & {15618} & {17.9} & {0.31} & {-} \\
    & {{*}2423} & {00 42 21.1, -09 54 29} & {\nodata} & {19.4} & {0.12}     &  {-}\\
\hline   
{13}& {322} & {00 42 18.7, -09 15 28} & {16,732} & {16.6} & {0.41}   &  {-}\\
    & {338} & {00 42 24.2, -09 16 16} & {18,195} & {17.1} & {0.25}   &  {-}\\
\hline   
{14}& {347} & {00 42 29.5, -10 01 07} & {15,165} & {17.7} & {0.29}   &  {-}\\
\hline
{15}& {374} & {00 42 41.5, -08 56 49} & {15,106} & {16.5} & {0.56}   &  {-}\\
    & {377} & {00 42 42.2, -08 55 28} & {16,992} & {16.6} & {0.43}   &  {-}\\
    & {385} & {00 42 44.2, -08 56 12} & {16,150} & {16.6} & {0.31}   &  {-}\\
\hline
{16}& {366} & {00 42 37.0, -09 45 20} & {17,065} & {17.8r} & {0.21}    &  {-}\\
    & {372} & {00 42 40.2, -09 44 17} & {16,922} & {17.8r} & {0.31}   &  {S0}\\
    & {382} & {00 42 43.9, -09 44 21} & {15,231} & {17.1} & {0.39}   &  {Sb}\\
\hline
{17}& {{*}2746} & {00 42 48.1, -09 34 54} & {\nodata} & {19.2} & {0.18}     &  {-}\\
    & {391} & {00 42 48.4, -09 34 41} & {17,940} & {17.7} & {0.33}   &  {-}\\
\hline
{18}    & {426} & {00 43 02.0, -09 46 40} & {14,734} & {17.3} & {0.18}   &  {-}\\
\hline
{19} & {423} & {00 43 01.4, -09 51 31} & {15,333} & {17.0} & {0.41}  &  {S0}\\
     & {{*}2923} & {00 43 04.9, -09 51 38} & {\nodata} & {20.1} & {0.08}    &  {-}\\
     & {{*}2934} & {00 43 05.0, -09 51 11} & {\nodata} & {18.8} & {0.13}    &  {-}\\
     & {435} & {00 43 06.0, -09 50 15} & {14,742} & {16.7} & {0.37}  &  {Sb}\\
     & {{*}2950} & {00 43 06.4, -09 51 40} & {17,727} & {18.2} & {0.24}&  {-}\\
     & {439} & {00 43 08.2, -09 49 37} & {15,203} & {17.2} & {0.22}  &  {-}\\
     & {442} & {00 43 10.1, -09 51 41} & {15,142} & {15.3} & {0.68}  &  {E/S0}\\
\hline
{20} & {447} & {00 43 10.9, -09 40 53} & {16,492} & {15.8} & {0.51}  &  {E}\\
     & {451} & {00 43 11.6, -09 38 16} & {16,253} & {15.8} & {0.48}  &  {Sb}\\
\hline    
{21} & {461} & {00 43 14.3, -09 10 21} & {15,015} & {18.4} & {0.28}&  {-}\\
\hline
{22} & {3114} & {00 43 19.5, -09 09 13} & {15,060} & {19.2} & {0.13}&  {-}\\
\hline
{23} & {486} & {00 43 31.2, -09 51 48} & {16,619} & {16.8} & {0.52} &  {S}\\
     & {{*}3234} & {00 43 32.6, -09 51 52} & {\nodata} & {19.7} & {0.12}   &  {-}\\
\hline 
{24} & {491} & {00 43 34.0, -08 50 37} & {14,968} & {17.0} & {0.33} &  {-}\\
     & {{*}3260} & {00 43 35.1, -08 51 13} & {\nodata} & {19.0} & {0.17}   &  {-}\\
\hline
{25} & {{*}3270} & {00 43 35.1, -09 32 14} & {\nodata} & {19.4} & {0.14}  &  {-}\\
     & {496} & {00 43 38.7, -09 31 21} & {15,004} & {17.0} & {0.34} &  {-}\\
\hline
{26} & {502} & {00 43 43.9, -09 04 23} & {15,004} & {16.8} & {0.40} &  {-}\\

\enddata
\tablecomments{Optical data obtained from the NED database (http://ned.ipac.caltech.edu).
Columns (1) and (2): ID for the field and the galaxies, respectively, using the
same names and references of Table \ref{fld}.
Column (3): R.A., Dec for each galaxy. Column (4): Optical radial velocity. 
Column (5): g-magnitude from NED; otherwise the band is indicated.
Column (6): Major angular diameter, in arcmins.
Column (7) Morphological type, when available. cD* : This object is wrongly classified;
it is a spiral object (see Sect. \ref{selected_fields})}
\end{deluxetable}


\begin{deluxetable}{ccccccc}
\tabletypesize{\scriptsize}
\tablecaption{The NIR magnitudes of observed galaxies in A\,85.\label{magn}}
\tablewidth{0pt}
\tablehead{
\colhead{ID} & \colhead{$J$} &\colhead{$H$} & \colhead{$K'$} 
&\colhead{$J_{2MASS}$}& \colhead{$H_{2MASS}$}& \colhead{$K_{2MASS}$}\\
\colhead{(1)} & \colhead{(2)} & \colhead{(3)} & \colhead{(4)} 
& \colhead{(5)} & \colhead{(6)} & \colhead{(7)}
}
\startdata

{{}145} & {14.54} & {16.09} &  {16.07} & {-} & {-} & {-}\\
{{}150} & {14.65} & {14.37} &  {13.41} & {-} & {-} & {-}\\

{{}167} & {14.05} & {13.57} &  {13.07} & {14.099} & {13.382} & {13.024}\\
{{*}1645} & {14.77} & {14.15} &  {13.79} & {14.866} & {14.085} & {13.888}\\
{{}174} & {13.47} & {13.07} &  {12.54} & {13.549} & {12.794} & {12.531}\\
{{}176} & {13.44} & {12.69} &  {12.48} & {13.397} & {12.720} & {12.384}\\
{{}177} & {13.14} & {12.56} &  {12.15} & {13.212} & {12.468} & {12.145}\\
{{}186} & {13.96} & {13.20} &  {12.76} & {14.039} & {13.300} & {12.977}\\

{{}175} & {14.61} & {14.02} &  {13.71} & {-} & {-} & {-}\\
{{}182} & {13.78} & {13.17} &  {12.92} & {13.820} & {13.156} & {12.789}\\
{{}192} & {13.99} & {13.43} &  {12.85} & {14.162} & {13.450} & {13.116}\\
{{}195} & {14.88} & {14.42} &  {14.03} & {15.083} & {14.619} & {13.866}\\
{{}197} & {13.19} & {12.69} &  {12.33} & {13.409} & {12.784} & {12.410}\\

{{}193} & {15.28} & {14.34} &  {14.05} & {15.147} & {14.356} & {13.707}\\
{{}201} & {14.43} & {13.94} &  {13.83} & {14.551} & {14.084} & {13.475}\\

{{}206} & {15.83} & {15.45} &  {14.92} & {-} & {-} & {-}\\
{{}209} & {14.37} & {13.55} &  {13.59} & {14.388} & {13.767} & {13.299}\\

{{}202} & {14.39} & {13.70} &  {13.36} & {14.273} & {13.573} & {13.132}\\
{{}210} & {13.89} & {13.17} &  {12.83} & {13.767} & {13.089} & {12.738}\\
{{}214} & {14.07} & {13.44} &  {13.36} & {14.110} & {13.413} & {13.202}\\

{{}215} & {14.98} & {15.51} &  {14.32} & {-} & {-} & {-}\\
{{}221} & {13.21} & {12.42} &  {12.17} & {13.188} & {12.512} & {12.092}\\
{{}222} & {15.41} & {16.18} &  {14.52} & {-} & {-} & {-}\\
{{}243} & {13.16} & {12.44} &  {12.02} & {13.169} & {12.418} & {12.070}\\

{{*}1895} & {15.22} & {14.64} &  {14.92} & {-} & {-} & {-}\\
{{}236} & {13.34} & {12.59} &  {12.53} & {13.394} & {12.644} & {12.333}\\
{{*}1951} & {14.94} & {14.30} &  {14.30} & {-} & {-} & {-}\\
{{}242} & {12.96} & {12.24} &  {11.99} & {12.856} & {12.082} & {11.741}\\
{{*}1966} & {16.66} & {16.29} &  {16.61} & {-} & {-} & {-}\\

{{}238} & {14.91} & {14.48} &  {14.87} & {-} & {-} & {-}\\
{{}251} & {13.12} & {12.49} &  {12.19} & {13.199} & {12.465} & {12.146}\\
{{}254} & {14.93} & {14.60} &  {13.92} & {14.986} & {14.428} & {14.054}\\
{{}255} & {14.74} & {14.15} &  {14.21} & {-} & {-} & {-}\\
{{}257} & {13.56} & {12.91} &  {12.75} & {13.634} & {12.885} & {12.587}\\

{{}276} & {14.12} & {13.26} &  {12.74} & {14.110} & {13.304} & {12.986}\\
{{}278} & {15.23} & {14.34} &  {13.97} & {15.172} & {14.347} & {13.993}\\

{{}286} & {14.24} & {13.39} &  {13.39} & {14.172} & {13.535} & {13.219}\\
{{*}2260} & {16.16} & {17.92} &  {18.35} & {-} & {-} & {-}\\

{{}315} & {15.45} & {14.50} &  {14.21} & {15.521} & {14.532} & {14.013}\\
{{*}2423} & {15.97} & {16.04} &  {14.97} & {-} & {-} & {-}\\

{{}322} & {13.88} & {13.12} &  {12.69} & {13.828} & {13.107} & {12.797}\\
{{}338} & {14.76} & {14.05} &  {13.83} & {14.681} & {14.026} & {13.533}\\

{{}347} & {16.32} & {15.10} &  {15.70} & {-} & {-} & {-}\\

{{}374} & {14.32} & {13.53} &  {13.72} & {14.353} & {13.600} & {13.396}\\
{{}377} & {13.92} & {13.14} &  {13.25} & {14.021} & {13.392} & {13.117}\\
{{}385} & {14.02} & {13.26} &  {13.01} & {14.059} & {13.256} & {13.068}\\

{{}366} & {16.82} & {15.68} &  {14.59} & {-} & {-} & {-}\\
{{}372} & {14.75} & {14.01} &  {13.93} & {14.898} & {14.148} & {13.807}\\
{{}382} & {14.53} & {13.84} &  {13.68} & {14.531} & {13.765} & {13.490}\\

{{*}2746} & {15.71} & {15.01} &  {14.64} & {-} & {-} & {-}\\
{{}391} & {15.80} & {15.41} &  {15.06} & {-} & {-} & {-}\\

{{}426} & {14.51} & {13.89} &  {13.66} & {14.661} & {13.858} & {13.698}\\

{{}423} & {14.55} & {13.60} &  {13.05} & {14.353} & {13.608} & {13.367}\\
{{*}2923} & {16.77} & {15.92} &  {15.43} & {-} & {-} & {-}\\
{{*}2934} & {16.48} & {15.71} &  {15.30} & {-} & {-} & {-}\\
{{}435} & {14.08} & {13.44} &  {12.97} & {14.067} & {13.281} & {13.014}\\
{{*}2950} & {16.60} & {16.03} &  {15.56} & {-} & {-} & {-}\\
{{}439} & {14.59} & {13.93} &  {13.43} & {14.495} & {13.808} & {13.380}\\
{{}442} & {13.02} & {12.06} &  {12.03} & {12.949} & {12.233} & {11.956}\\

{{}447} & {13.53} & {12.92} &  {12.84} & {13.571} & {12.877} & {12.572}\\
{{}451} & {13.91} & {13.27} &  {13.08} & {13.804} & {13.084} & {12.778}\\

{{}486} & {15.68} & {14.97} &  {14.77} & {-} & {-} & {-}\\
{{*}3234} & {17.83} & {16.21} &  {16.17} & {-} & {-} & {-}\\
{{}491} & {15.65} & {15.24} &  {14.84} & {-} & {-} & {-}\\
{{*}3260} & {16.38} & {16.09} &  {16.14} & {-} & {-} & {-}\\

{{*}3270} & {16.88} & {15.77} & {15.48} & {-} & {-} & {-}\\
{{}496} & {15.91} & {14.88} &  {15.03} & {-} & {-} & {-}\\

{{}502} & {16.31} & {15.59} &  {15.21} & {-} & {-} & {-}\\

\enddata
\tablecomments{Column (1): Galaxy names, as in Table \ref{opt}.
Columns (2), (3), (4): NIR (J,H,K'), 14\prin\, aperture magnitudes obtained in the present work.
Columns (5), (6), (7): NIR (J,H,K'), 14\prin\, aperture magnitudes from 2MASS, when available,
for comparison.
}
\end{deluxetable}

\newpage

\begin{deluxetable}{ccccc}
\tabletypesize{\scriptsize}
\tablecaption{Asymmetry index for selected galaxies in A\,85.\label{asym}}
\tablewidth{0pt}
\tablehead{
\colhead{ID} & J-H  &\colhead{\atf} & 
\colhead{Position} &  \colhead{Dist(Mpc)}
\\

\colhead{(1)} & \colhead{(2)} & \colhead{(3)} & \colhead{(4)} 
& \colhead{(5)}
}\startdata 
{150} & {0.53} &{0.032}& {C2} & {0.62}  \\
{167} & {0.44} &{0.070}& {C2} & {0.49}  \\
{174} & {0.52} &{0.010}& {C2} & {0.41}  \\
{175} & {0.60} &{0.050}& {M}  & {0.35}  \\
{176} & {0.64} &{0.140}& {C2} & {0.33}  \\
{177} & {0.82} &{0.110}& {M}  & {0.37}  \\
{182} & {0.61} &{0.049}& {C2} & {0.29}  \\
{186} & {0.65} &{0.070}& {M}  & {0.32}  \\
{193} & {0.85} &{0.050}& {}   & {1.19}  \\
{197} & {0.68} &{0.250}& {C2} & {0.31}  \\
{201} & {0.79} &{0.150}& {}   & {1.26}  \\
{209} & {0.78} &{0.040}& {SB} & {0.58}  \\
{214} & {0.59} &{0.030}& {C2} & {0.14}  \\
{221} & {0.72} &{0.045}& {SB} & {0.56}  \\
{236} & {0.76} &{0.050}& {M}  & {0.07}  \\
{242} & {0.64} &{0.062}& {M}  & {0.0}  \\
{243} & {0.69} &{0.059}& {SB} & {0.45}  \\
{251} & {0.65} &{0.140}& {SB} & {0.79}  \\
{255} & {0.56} &{0.037}& {SB} & {0.77}  \\
{257} & {0.64} &{0.320}& {SB} & {0.78}  \\
{276} & {0.79} &{0.110}& {}   & {1.80}  \\
{278} & {0.82} &{0.100}& {}   & {1.95}  \\
{286} & {0.79} &{0.030}& {SB} & {0.96}  \\
{322} & {0.73} &{0.023}& {M}  & {0.45}  \\
{338} & {0.72} &{0.017}& {M}  & {0.58}  \\
{347} & {0.65} &{0.040}& {}   & {2.93}  \\
{372} & {0.71} &{0.100}& {SE} & {1.80}  \\
{374} & {0.65} &{0.020}& {}   & {1.66}  \\
{377} & {0.63} &{0.040}& {}   & {1.20}  \\
{382} & {0.65} &{0.190}& {SE} & {1.96}  \\
{385} & {0.69} &{0.020}& {}   & {1.21}  \\
{391} & {0.50} &{0.030}& {}   & {1.46}  \\
{423} & {0.72} &{0.020}& {SE} & {1.80}  \\
{435} & {0.73} &{0.025}& {SE} & {2.48}  \\
{442} & {0.84} &{0.041}& {SE} & {2.50}  \\
{447} & {0.68} &{0.060}& {SE} & {2.02}  \\
{451} & {0.67} &{0.004}& {SE} & {1.90}  \\
{486} & {0.69} &{0.030}& {SE} & {2.79}  \\
{491} & {0.54} &{0.040}& {}   & {2.52}  \\
{496} & {0.50} &{0.010}& {}   & {2.00}  \\
{502} & {0.59} &{0.010}& {}   & {2.10}  \\
\enddata

\tablecomments{Asymmetry index for the galaxies being larger
than 0.25\prim.
Column (1): galaxy name, as in previous tables.
Column (2): (J-H) color index.
Column (3): the \atf\, index, after applying the $bmodel$ residual.
Column (4): the projected position of each galaxy, following the code used
in (BA09) for the substructures reported across A\,85. Galaxies for which
a $Position$ is not given, are considered at the outskirts of the cluster.
Column (5): projected cluster-centric distance, in Mpc.  }
\end{deluxetable}


\end{document}